%% file: multvel_lda_v7_arxiv.tex
\documentclass[12pt]{aastex}
\usepackage{xspace}
\usepackage{graphicx}
\usepackage{natbib}
\usepackage{amssymb}
\newcommand{\nraoblurb}{The National Radio Astronomy Observatory is
a facility of the National Science Foundation operated under cooperative
agreement by Associated Universities, Inc.}
\newcommand{\hide}[1]{}

\newcommand{\lsim}{\ensuremath{\,\lesssim\,}\xspace}
\newcommand{\gl}{\ensuremath{\ell}\xspace}
\newcommand{\gb}{\ensuremath{{\it b}}\xspace}

\newcommand{\lb}{\ensuremath{(\gl,\gb)}\xspace}

\newcommand{\kms}{\ensuremath{\,{\rm km\,s^{-1}}}\xspace}

\newcommand{\pc}{\ensuremath{\,{\rm pc}}\xspace}
\newcommand{\kpc}{\ensuremath{\,{\rm kpc}}\xspace}

\newcommand{\mK}{\ensuremath{\,{\rm mK}}\xspace}

\newcommand{\degree}{\ensuremath{\,^\circ}\xspace}

\newcommand{\te}{\ensuremath{{T_{e}}}\xspace}

\newcommand{\hi}{{\rm H\,{\footnotesize I}}\xspace}
\newcommand{\hii}{{\rm H\,{\footnotesize II}}\xspace}

\newcommand{\hal}[1]{\ensuremath{{\rm H}\,{#1}\,\alpha\xspace}}
\newcommand{\hna}{\ensuremath{{\rm H}\,{\rm n}\,\alpha\xspace}}

\newcommand{\cor}{\ensuremath{^{\rm 13}{\rm CO}}\xspace}

\begin{document}

\title{Untangling the Recombination Line Emission from HII Regions with Multiple Velocity Components}

\author{L.~D.~Anderson\altaffilmark{1,2}, L.~A.~Hough\altaffilmark{1},
T.~V.~Wenger\altaffilmark{3}, T.~M.~Bania\altaffilmark{4}, Dana~S.~Balser\altaffilmark{5}}

\altaffiltext{1}{Department of Physics and Astronomy, West Virginia
  University, PO Box 6315, Morgantown, WV 26506, USA}
\altaffiltext{2}{Adjunct Astronomer at the National Radio
  Astronomy Observatory, P.O. Box 2, Green Bank, WV 24944, USA}
\altaffiltext{3}{Department of Astronomy, University of Virginia,
  P.O. Box 3813, Charlottesville, VA 22904, USA}
\altaffiltext{4}{Institute for Astrophysical Research, Department of
  Astronomy, Boston University, 725 Commonwealth Avenue, Boston, MA
  02215, USA}
\altaffiltext{5}{National Radio Astronomy Observatory,
  520 Edgemont Road, Charlottesville VA, 22903-2475, USA}

\begin{abstract}
\hii\ regions are the ionized spheres surrounding high-mass stars.
They are ideal targets for tracing Galactic structure because they are
predominantly found in spiral arms and have high luminosities at
infrared and radio wavelengths.  In the Green Bank Telescope
\hii\ Region Discovery Survey (GBT HRDS) we found that $>30\%$ of
first Galactic quadrant \hii\ regions have multiple hydrogen radio
recombination line (RRL) velocities, which makes determining their
Galactic locations and physical properties impossible.  Here we make
additional GBT RRL observations to determine the discrete \hii\ region velocity for all
\input n multiple-velocity sources within
$18\degree<\ell<65\degree$.  The multiple-velocity sources are
concentrated in the zone $22\degree < \ell < 32\degree$, coinciding
with the largest regions of massive star formation, which implies that
the diffuse emission is caused by leaked ionizing photons.  We combine our observations with analyses of the electron
temperature, molecular gas, and carbon
recombination lines to determine the source velocities for \input
resolved discrete \hii\ regions ($\input percentresolved\%$ of the sample).
With the source velocities known, we resolve the kinematic distance
ambiguity for \input kda_resolved regions, and thus determine their
heliocentric distances.
\end{abstract}

\keywords{Galaxy: structure -- H II regions -- infrared: ISM -- stars: formation}

\section{Introduction\label{sec:intro}}
\hii\ regions are created by the ionizing radiation from massive
stars.  Only stars with spectral types of B0 or earlier are capable of
producing the ultra-violet photons necessary to appreciably ionize the
surrounding interstellar medium (ISM).  Such stars only live
$\sim10$\,Myr and therefore \hii\ regions trace star formation at the
present epoch.  Since \hii\ regions surround massive stars, they trace
spiral structure and can be used to understand the structure of our
Galaxy.  \hii\ region distances are required to turn measured
properties (i.e., flux and angular size) into physical properties
(i.e., luminosity and physical size).  Distances are also essential
for using \hii\ regions for studies of Galactic structure and
large-scale Galactic star formation.

In the Green Bank Telescope \hii Region Discovery Survey \citep[GBT
  HRDS;][]{bania10, anderson11}, we detected radio recombination line
(RRL) emission from 448 previously unknown Galactic \hii\ regions at
X-band (9\,GHz; 3\,cm).  Of the 448 detected targets, 130 ($\sim
30\%$) have multiple RRL velocity components: 106 have two components,
23 have three components, and one has four components.  Without
knowing the source velocity, it is not possible to compute kinematic
distances or to derive physical properties (including electron
temperature, e.g., Balser et al., 2015, submitted) for the multiple-velocity
HRDS \hii\ regions.

\hii\ regions with multiple RRL components are common in the
literature, but their numbers have increased with the greater
sensitivity of modern instruments.  Of the 462 \hii\ regions detected
in RRL emission by \citet{lockman89}, 17 ($<\!4\%$) have multiple
components.  In the more recent RRL surveys of ultra-compact
\hii\ regions by \citet{araya02, watson03, sewilo04b}, the combined
percentage of multiple-component RRL \hii\ regions detected is 12\%.
That nearly 30\% of HRDS sources have multiple line components speaks
to the sensitivity of the GBT and its ACS spectrometer, and to our
efficient observational setup.  Within the 2\,GHz bandpass at X-band,
we were able to simultaneously measure the emission from seven
hydrogen RRLs in two polarizations.  This allowed very sensitive
measurements in short observations; the average on-target integration
was $\sim 10$\,min.



There are at least two explanations for the origin of the multiple RRL
velocity components.  In previous RRL work on \hii\ regions, most
authors have assumed that each velocity component arises from a
discrete \hii\ region along the line of sight.  By ``discrete''
\hii\ region we mean RRL and free-free continuum radiation stemming
from locations that are spatially well-defined on the sky. This
emission is distinct from that emanating from diffuse ionized gas that
produces RRLs, but does not have detectable continuum emission above
the background.  

The spatial density of discrete \hii\ regions is not high enough
  to explain the HRDS multiple-velocity detections, especially given
  the small GBT beam at X-band ($82\arcsec$).  Using the WISE Catalog
  of Galactic \hii\ regions \citep{anderson14}, we estimate the
  fractional sky area covered by \hii\ regions.  The WISE catalog
  contains all known Galactic \hii\ regions, and lists approximate
  \hii\ region sizes based on their WISE mid-infrared emission.  The
  highest density of multiple-velocity \hii\ regions is found in the
  zone $\ell \simeq 22-32\arcdeg$, $|b| \le 1\degree$ (see below).
  Using the WISE catalog sizes, we find that only 16\% of the sky area
  in this zone is covered by known \hii\ regions with measured ionized
  gas spectral lines (including ``grouped'' \hii\ regions in large
  complexes). Just over 2\% of the sky area in this zone is covered by
  more than one region.  Even including \hii\ region candidates with
  weak expected RRL emission that would not have been detected in the
  HRDS, these numbers are just 22\% and 3\%, respectively.  Most of
  these overlap regions are in fact caused by \hii\ regions near the
  same velocity that therefore cannot be the cause of multiple RRL
  lines.  When computing the fractional sky areas, we factored in the
  GBT X-band beam size by counting regions closer than $82\arcsec$ as
  overlapping.  For an infinitely small beam these numbers would
  further decrease.  Therefore, it seems highly unlikely that the
  multiple-velocity \hii\ regions can be caused by confusion along the
  line of sight from \hii\ regions at different velocities.
  Furthermore, the spatial morphology of the multiple-component
  \hii\ region images at infrared and radio wavelengths is the same as
  that of the single-component HRDS sources.  There are no additional
  radio continuum sources seen in the HRDS data, and no additional
  WISE sources seen superposed on top of their mid-infrared emission.
  There is no indication that there are two distinct \hii\ regions
  along the line of sight.  This suggests that some of the additional
  velocity components are due to diffuse ionized gas; these components
  would thus stem from the ``warm ionized medium'' (WIM).

The $10^4$\,K WIM contains up to 90\% of the ionized mass in the
Galaxy \citep{haffner09} and is therefore an important, but poorly
understood, component of the ISM.  It has alternatively been called
the ``Diffuse Ionized Medium,'' the ``Galactic Ridge Emission,'' or
the ``Extended Low-Density Medium,'' albeit sometimes with slightly
different definitions.  It has long been thought that massive
stars produce the ionizing photon fluxes necessary to create and
sustain the WIM \citep[e.g.,][]{reynolds84, domgorgen94}.  The WIM
detected at radio wavelengths is composed of both ``worms'' and
``chimneys'' blown out by supernova remnants or successive generations
of O-stars \citep[e.g.,][]{heiles96}, and also of extended diffuse
emission from large star-forming regions \citep[see][]{murray10,
  roshi12}.

Recent results from \citet{murray10} have confirmed what has been
known for some time: \hii\ regions are leaky and a large number of the
ionizing photons escape beyond their photo-dissociation regions
\citep[see][]{oey97}.  A recent study by \citet{anderson15} showed
that even for the ``perfect bubble'' \hii\ region RCW\,120, where the
photodissociation region is thick and extends completely around the
\hii\ region, $\sim 25\%$ of the ionizing photons are leaking into the
nearby ISM.

We hypothesize that most of the additional velocity components are due
to the WIM.  Lending support to this hypothesis, the \hii\ regions
with multiple velocity components are not evenly distributed in the
Galaxy, but are clustered near the locations of large star-forming
regions that may be leaking photons into the WIM.  For example,
\citet{anderson11} show that 21 of 23 \hii\ regions within $1\degree$
of W43 at \lb=($31.78\degree, -0.03\degree$) have multiple velocity
components.  Of these 21 \hii\ regions, all have one velocity
component within 10\,\kms\ of the velocity of W43,
91.6\,\kms\ \citep{lockman89}.  Furthermore, the strength of the
velocity component within 10\,\kms\ of the velocity of W43 decreases
with distance from this nominal central position of W43.  This
is consistent with the idea that the WIM is created and maintained by
leaking photons from large \hii\ regions.

No study has yet proven that the large-scale distribution of the WIM
is connected to the emission from the Galactic population of discrete
\hii\ regions.  In part this is because the WIM has traditionally been
studied at optical wavelengths, and most Galactic \hii\ regions are
not optically visible.  In radio observations of the WIM,
\citet{heiles96} showed that there were plumes of ionized gas
extending above and below the plane, and that these plumes were in
some cases related to massive star formation regions in the Galactic
plane.  \citet{alves10, alves12, alves14} discussed
low-resolution ($\sim 15\arcmin$) studies of the RRL emission in the
Galactic plane.  At the 1.4\,GHz frequency of their observations, the
intensity of the emission from low-density plasma is enhanced compared
to that of observations at higher frequencies.  With their low spatial
resolution data, however, it is not possible to disentangle the
emission from the diffuse ionized gas and discrete \hii\ regions, or to
investigate the connection between them.

Here we make RRL observations toward our HRDS sample of first
Galactic quadrant \hii\ regions with multiple RRL velocities.  Our goal is to
resolve which velocity component arises from the
discrete \hii\ region.

\section{Observations and Data Analysis}
We used the GBT 100\,m telescope to observe RRLs in a sample of
multiple RRL velocity \hii\ regions from 2011 December through 2013
April.  Our spectral line observational setup was identical to that of
the HRDS \citep[described in][]{anderson11}.  Our observations made
total-power position-switched 6\,min. on-target and 6\,min. off-target
integrations at X-band (9\,GHz; 3\,cm), hereafter called ``pairs.''
The on- and off-target integrations followed the same path on the sky,
and we observed a single pair for each position.  We simultaneously
observed the \hal{87} to \hal{93} transitions.  We began each
observation by measuring a source of known position and intensity in
order to obtain pointing and focus corrections.  We made no
corrections for atmospheric effects.

We reduce the data with the TMBIDL software (T.\,M.~Bania, 2015,
private communication). We average the seven transitions together to
improve the RRL signal-to-noise ratio \citep{balser06} and smooth this
average \hna\ spectrum to a velocity resolution of 1.86\kms.  This
resolution is more than sufficient to resolve the typical
25\,\kms\ full width-half maximum (FWHM) of hydrogen RRLs
\citep{anderson11}.  We remove a baseline (typically third-order) and
fit a Gaussian model to each detected RRL component.  From the
Gaussian fits we derive the line intensity, LSR velocity, and FWHM for
each component.

Our source sample consists of all \hii\ regions located within $\ell =
18\degree$ to $65\degree$ whose RRL spectra have multiple detected
velocity components.  We restricted the current study to this
longitude range because of the availability of \hi\ data (used in
Section~\ref{sec:distances}) and the relative accuracy of kinematic
distances \citep[see][Wenger et al., 2015, ins prep.]{anderson12c}.
There are \input n multiple-velocity \hii\ regions within this range:
\input nliterature from the literature and \input nhrds from the HRDS.
The literature sources were first observed in RRL emission near 5\,GHz
by \citet{lockman89}, \citet{lockman96}, or \citet{sewilo04b}.
Figure~\ref{fig:multvel_pos} shows that these multiple-velocity
\hii\ regions are not evenly distributed across the Galaxy, but rather
are found preferentially in the zone $\ell \simeq 22-32\arcdeg$,
coinciding with active regions of high mass star formation in the
inner Galaxy.

We conduct our observations in two phases.  We first observe the
\input nliterature literature sources on-target at the nominal peak of
the free-free radio continuum intensity to establish X-band (3\,cm)
line parameters and to verify the measured velocities.  We fit
Gaussians to each detected line component.  In
Table~\ref{tab:literature} we give the derived RRL peak antenna
temperature, LSR velocity, FWHM, and rms noise.  The \input
nliterature on-target observations resulted in \input nliteraturetotal
detected lines: 5 observations have one component, 21 have two
components, and two have three components.  When multiple RRL
components are detected, we follow our usual convention and append
``a'' to the source name of the strongest component, ``b'' to the
source name of the next strongest component, etc.  We did not detect
multiple velocity components for \input reason1 of these literature
\hii\ regions, which could be caused by either a lack of sensitivity
in our observations or an error in the original observations.

We next observe, using the same observational setup, positions near
all multiple-velocity \hii\ regions (literature and HRDS).  The
goal of these observations is to identify the discrete \hii\ region
source velocities.  The WIM is more broadly distributed in the
Galactic disk compared to a discrete \hii\ region.  Our hypothesis is
that observations at positions offset from the multiple-velocity
\hii\ regions (``off-target'') will detect only the WIM.

The off-target positions are at least one $82\arcsec$ beam width away
from the multiple-velocity \hii\ regions and any other nearby radio
continuum sources.  A typical separation is a few arcminutes.  We
determine the off-target positions by examining VGPS 21\,cm continuum
data \citep{stil06} to identify locations devoid of discrete sources
of radio continuum emission (Figure~\ref{fig:offtarget}).  For the 51
cases where the results from the first off-target position do
  not definitively indicate which line is from the discrete
  \hii\ region, we observe a second off-target position. In 12 cases,
we need a third off-target position and in two cases we need a fourth.
Some off-target positions are nearly equidistant between two
multiple-velocity \hii\ regions, and single off-target observation can
be used for both multiple-velocity regions.  In total, we observe
\input noffpointing off-target positions for the \input
  nliterature literature and \input nhrds HRDS regions.

The observations of the \input noffpointing off-target positions
resulted in the detection of \input noff hydrogen RRLs, or just fewer
than two velocity components per position.  We give the derived
off-target line parameters in Table~\ref{tab:offtarget}, which lists
the Galactic longitude and latitude, the separation from the nearest
multiple-velocity \hii\ region, the line intensity, the LSR velocity,
the line FWHM, and the rms noise.  The final column gives the
multiple-velocity \hii\ region(s) that use this off-target position in
our subsequent analyses.  Line parameter uncertainties in
Table~\ref{tab:offtarget} are $1\sigma$.

\section{Determining Discrete HII Region Velocities and Kinematic Distances\label{sec:analysis}}

\subsection{Discrete HII Region Velocity Criteria}
Our goal is to use these observations to identify the velocity of the
discrete \hii\ regions.  To do this we create a database of \input
ntotal X-band RRL line parameters from $\ell = 18\degree-65\degree$:
\input nliteraturetotal lines from the \input nliterature on-target
observations of literature \hii\ regions (Table~\ref{tab:literature}),
\input noff lines from the \input noffpointing off-target positions
(Table~\ref{tab:offtarget}), and \input nhrdstotal lines from the
\input nhrds multiple-velocity HRDS sources \citep{anderson11}.  In
order to determine which velocity components are from the discrete
\hii\ regions, we use pairs of on- and off-target spectra from this
database to search for velocity components with substantially
different intensities.

To search for lines that decrease in intensity at the off-target
locations, we must associate on-target and off-target velocity
components.  Due to motions within the ISM, the fact that we are not
sampling the same column of diffuse plasma, and measurement error, we
cannot expect the line velocities of the on-target and off-target RRL
parameters to be identical.  We associate a velocity component from an
on-target spectrum with a component from an off-target spectrum if
their velocities are within 10\,\kms\ of each other.
Sometimes there are two components in an off-target spectrum that are
both within 10\,\kms\ of a single component in an on-target spectrum.
In such cases, we associate both of these components with the single
on-target component.

We examine six independent criteria to determine the discrete
  \hii\ region velocities:\\
\begin{enumerate}
\item We check if only one line was
detected in our on-target observations of the literature sources. As
mentioned previously, this may be caused by incorrect line parameters
in the literature or by different beam areas for the two sets of
observations.  For these sources the discrete \hii\ region velocity is unambiguous.
\item We identify HRDS sources that have a negative velocity for one
  component.  For the longitude range of our sources, a negative RRL
  velocity requires the emitting plasma to be outside the Solar
  circle.  Models of ionized gas in the Galaxy
    \citep[e.g.,][]{taylor93} show a very low density outside the Solar
    circle.  Our observations probably lack the sensitivity to detect
    such low density ionized gas.  Detected negative velocity RRLs in
    the first Galactic quadrant are therefore likely to be from discrete
    \hii\ regions.
\item We examine the
difference in intensity between the on- and off-target observations for each associated velocity component.
A large decrease in line intensity off-target for one 
component is a clear indication that this velocity is associated with the discrete
\hii\ region (explained in detail below).  
\item We use the derived
electron temperatures for each RRL component to determine which on-target
  line results in an electron temperature within the range found in
  previous studies of Galactic \hii\ regions (explained in detail
below).  
\item We use the association of each on-target RRL component with
  molecular gas emission.  These molecular data come from two sources:
  the spectroscopic and morphological analysis of \hii\ regions
    in \cor\ data from the Galactic Ring Survey \citet{jackson06} by
    \citet{anderson09b}, and the svelocities compiled in the WISE
    catalog of Galactic \hii\ regions \citep[][Version
      1.3]{anderson14}.  Molecular gas is associated with massive star
    formation, but not with the WIM. If molecular gas is only
    associated with one of the velocity components, this is most
    likely the discrete \hii\ region velocity.
\item We examine whether a carbon RRL is associated with one of the
  RRL velocity components.  Carbon RRLs are created in the PDRs of
  \hii\ regions, and therefore the carbon RRLs should be strongest
  near the discrete \hii\ region velocity.  Observations of the WIM do
  contain carbon RRLs \citep[e.g.,][]{roshi02}, but to date these have
  only been detected at low frequencies ($\lsim 1$\,GHz).  We use
    the carbon RRL parameters for the HRDS sources from
    \citet{wenger13}; this analysis is not possible for the literature
    sources because there is no comparable carbon RRL catalog.
\end{enumerate}

The RRL component intensity difference between the on- and off-target
spectra (the third metric above) allows us to discriminate between the
discrete and diffuse velocity components.  In Figure~\ref{fig:spectra}
we show example spectra that have differences in line intensity
between the on- and off-target positions.  The magnitude of the
intensity decrease between on- and off-target RRL spectra must be
empirically set.  The brightest diffuse emission detected in our
observations is $\sim 50$\,mK (see below). Therefore, any line
that decreases in intensity by 50\,mK between on- and off-target
positions is likely to be from the discrete \hii\ region.  We find
that our derminations of the discrete \hii\ region velocity based on
line intensity differences of 20\,mK agree with those determined using
other criteria.  We therefore take decreases of 20\,mK to be a
reliable criterion, although we assume that it is less reliable than
decreases of 50\,mK.  Lines that decrease by 10\,mK, however, show
considerably worse agreement with other criteria, and therefore we do
not use 10\,mK as an intensity difference criterion.

We naively would not expect to detect large decreases in the diffuse
RRL intensity with small changes in position, and therefore the 50\,mK
criterion is rather conservative (which is why we included the 20\,mK
criterion as well).  Our experiment does show, however, that for some
sight lines multiple RRL component intensities decrease by up to
50\,mK.  This implies either that the diffuse ionized gas is quite
clumpy, that a single discrete region is causing both lines, or that
there are somehow two discrete \hii\ regions that cannot be separated
even in the high resolution MIR data.  Regardless of the reason that
both lines decrease significantly in intensity for some directions,
the 50\,mK criterion is justifiably conservative.  The median line
intensity in the HRDS was $\sim20$\,mK, and therefore the 20\,mK
criterion is also rather conservative.
 
Galactic \hii\ regions have a rather narrow range of electron
temperatures, \te, ranging from $\sim5,\,\!000$\,K to
$\sim10,\,\!000$\,K \citep[e.g.,][]{quireza06b}. The \te of a discrete
\hii\ region plasma should lie within this observed range.  We thus
can use the electron temperature (fourth metric above) to assess which
on-target RRL component gives the most reasonable \te\ value, and thus
determine the source velocity.

In local thermodynamic equilibrium (LTE), the electron temperature can be derived from observable quantities:
\begin{equation}
T_e =  7103.3 \left(\frac{\nu}{\rm GHz}\right)^{1.1} \left[\frac{T_C}{T_L({\rm H}^+)}\right] \left[\frac{\Delta v({\rm H}^+)}{\rm km\,s^{-1}}\right]^{-1} \left[ 1 + \frac{n(^4{\rm He}^+)}{n({\rm H}^+)} \right]^{-1}\,,
\end{equation}
where $\nu$ is the observing frequency, $T_C/T_L$ is the peak
continuum-to-line intensity ratio, $\Delta v$ is the Hn$\alpha$ RRL
line width, and $n(^4{\rm He}^+)/n({\rm H}^+)$ is the helium ionic
abundance ratio, $y+$.  Each multiple-velocity \hii\ region has a
single $T_C$ value, but multiple $T_L$ and line width values, leading
to a different \te\ for each RRL.  This allows us to determine
which lines produce \te\ values outside the nominal range.  Following
Balser et al., (2015, submitted), who used the same observational
configuration, we use 8.9\,GHz for $\nu$, and assume $y^+ = 0.07$
\citep{quireza06b}.  We must assume a value for the helium ionic
abundance ratio here because, for the integration times we used,
$^4$He$^+$ is too weak to measure for these sources.  In fact, for the
range of $y^+$ observed in Galactic \hii\ regions the electron
temperature does not depend strongly on this quantity
\citep{balser11}.

As for the line intensity metric, we must empirically
  determine a reasonable range of \te\ values.  We define this range
  using the single-velocity \hii\ region data from \citet{anderson11}.
  We further divide the single-velocity population into ``high
  quality'' and ``low quality'' to investigate the effects of data
  quality on the derived values of \te.  These high quality sources
  have a simple radio continuum profile \citep[complex flag equal to
    zero in][]{anderson11} and a continuum intensity of more than
  $100$\,mK.  We compare the single-velocity \te\ values with that of
  the multiple-velocity \hii\ regions, computed using the on-target
  RRL parameters from the present work and the measurements of $T_C$
  from \citet{anderson11}.

We show in the top panel of Figure~\ref{fig:te} the electron
temperature distribution for all HRDS sources.  The spread in electron
temperature values for multiple-velocity \hii\ regions is clearly
greater than that of the single-velocity \hii\ regions.  We expect the
derived electron temperature for the diffuse RRL components to be
overestimated in general since $T_e \propto T_C/T_L$ and the diffuse
line intensity is low.  No single-velocity HRDS source has a
\te\ value greater than $2\times 10^4$\,K.  The derived electron
temperatures for multiple-velocity \hii\ regions are on average over
twice as large as those of the single-velocity \hii\ regions.  The
averages for the high-quality single-velocity, low-quality
single-velocity, and multiple-velocity samples are \input
singletestats\,K, \input singlecomplextestats\,K, and \input
multipletestats\,K, respectively.

Because metals are the main coolants of the \hii\ region plasma, the
electron temperature gradient is a proxy for the metallicity 
  radial gradient \citep{shaver83}.  The electron temperature is the
lowest toward the Galactic center, where the ISM has been enriched by
multiple generations of stars (high metallicity), and is highest in
the outer reaches of the Milky Way, where there have been fewer
stellar generations (low metallicity).  
The electron temperature Galactocentric radial gradient is
  apparent in the bottom panel of Figure~\ref{fig:te}, which shows the
  electron temperature of HRDS sources within $18\degree < \ell <
  65\degree$ as a function of Galactocentric radius.  The solid
  curve is a fit to the high-quality single-velocity \hii\ regions
  (filled black points) of the form $T_e = a + b \, R_{Gal}$, where
$a = 2800\pm340$\,K and $b = 560\pm60$\,K\,kpc$^{-1}$.  It is
important to note that this fit characterizes the sample well enough
for our purposes, but is unreliable for the study of electron
temperature gradients themselves because we have not accounted for the
relatively poor data quality of some sources.  A complete treatment of
the electron temperature distribution of HRDS sources is given by
Balser et al. (2015, submitted) and Wenger et al. (2015, in prep.).

We define two zones of acceptable \te\ values that contain at least
90\% of all single-velocity \hii\ regions: one zone for high-quality
sources and one for low-quality.  Within the area defined by changes
to the y-intercept of $\pm6\sigma$ (dashed curves in bottom panel of
Figure~\ref{fig:te}; $a = 760$ and $4840$\,K.)  lie \input
singletepercentage $\!\!$\% of all high-quality single-velocity
sources.  This area defines acceptable electron temperatures for
high-quality sources.  Within an area below a change to the
y-intercept of $+10\sigma$ (dotted curve; $a = 6200$\,K) are \input
singletecomplexpercentage $\!\!$\% of all low-quality single-velocity
sources.  This area defines acceptable electron temperatures for
low-quality sources.  We do not use a change of $-10\sigma$ to the
y-intercept to define the area.  The large number of low-quality
single-velocity \hii\ regions with very low \te\ values (apparent in
both panels of Figure~\ref{fig:te}) indicates that this would not lead
to a reliable criterion.  For multiple-velocity \hii\ regions, lines
that imply \te\ values within these acceptable zones are likely from
discrete \hii\ regions, while those outside of the zones are more
likely to be from diffuse gas.

Using the above analyses, we refine the 6 metrics to develop the
criteria used for the determination of the discrete \hii\ region
velocities: 1) only one on-target velocity component was detected
(literature sources only); 2) one on-target velocity component has a
negative LSR velocity; 3a) the intensity of one off-target line
component is at least 50\,mK less than that of the on-target position;
3b) the intensity of one off-target line component is between 20\,mK
and 50\,mK less than that of the on-target position (the former being
the more reliable criterion); 4a) the electron temperature derived
from only one line component is within the reasonable range, for high
quality sources; 4b) the electron temperature derived from only one
line component is within the reasonable range, for low quality
sources; 5a) \citet{anderson09b} find CO emission matching the source
morphology at only at one velocity (their quality factors of ``A'' or
``B''); 5b) dense molecular gas, compiled in the WISE catalog of
Galactic \hii\ regions \citep{anderson14}, is only found at one
velocity; 6) a carbon RRL is found at only one of the RRL velocities.

These criteria allow us to identify the discrete \hii\ region velocity
component for \input resolved of the \input n multiple-velocity
\hii\ regions (\input percentresolved $\!\!$\%).  If two criteria give
conflicting results, we do not assign any velocity; this applies to
only 2 sources.  For the sources for which we observed multiple
  off-target positions, we require that most positions give the same
  answer (i.e., both if there are two off-target position or two if
  there are three). We summarize the efficacy of our criteria in
Table~\ref{tab:criteria}, which lists the number of sources that the
criterion applies to, and the percentage of all sources that this
represents.  Since multiple criteria can be used for a given source,
the percentages do not add to 100\%.  Table~\ref{tab:criteria} shows
that the most useful criteria are from the off-target observations
(54\% of all sources), and the next most useful are from the electron
temperature analysis (46\% of all sources).  We use multiple criteria
for $48\%$ of all \hii\ regions, and 58\% of all determinations.
Because we require that all criteria be in agreement in order to
determine the source velocity, and because we have no objective
measure of reliability, we cannot evaluate the accuracy of the
individual criteria.

In Figure~\ref{fig:height} we show that the discrete \hii\ region RRL
intensities are greater on average than those of the more diffuse gas.
The average peak line intensity for the discrete \hii\ regions is
59.7\,mK (with a large dispersion), whereas it is 14.9\,mK for the
more diffuse gas.  For 91 of the \input resolved regions (88\%), the
discrete \hii\ region velocity is the brightest of the multiple lines
detected.  As expected, the distributions of on- and off-target
diffuse line parameter intensities are similar.  The vertical lines in
Figure~\ref{fig:height} show the intensity criteria of 20\,mK and
50\,mK.  The fact that the diffuse line intensities are generally
lower than these values lends some support to these choices of
intensity criteria.  More than 98\% of the diffuse line intensities
are less than 50\,mK, and 80\% are less than 20\,mK.

\subsection{Kinematic Distances \label{sec:distances}}
Using the derived \hii\ region velocities we can compute kinematic
distances using a Galactic rotation curve model.  Here, we use the
\citet{brand93} rotation curve. In the inner Galaxy there are two
possible kinematic distances for each positive velocity, a problem
known as the kinematic distance ambiguity (KDA).  We make a kinematic
distance ambiguity resolution (KDAR) using \hi\ absorption, in the
same manner as \citet{anderson09a} and \citet{anderson12c}.  To
resolve the KDA we examine \hi\ on- and off-source spectra.  The
difference of these two spectra shows absorption of background 21\,cm
radio continuum emission from the \hii\ region by foreground \hi.  If
absorption only occurs up to the \hii\ region velocity, the source
must lie at the near distance.  If absorption is detected between the
\hii\ region velocity and the tangent point velocity, the source must
lie at the far distance.  If a source velocity is within 10\kms of the
tangent point velocity, we use the tangent point distance.
As in previous work, the
off-target spectrum was taken at a position as close to the source as
possible so we can sample a similar column of hydrogen.  We assign
each {KDAR a quality factor based on our qualitative assessment of our
  confidence in the KDAR: ``A'' is very certain, ``B'' is less
  certain, and ``C'' means we cannot resolve the KDAR.
Of the \input kda_possible inner Galaxy \hii\ regions for which we
know the source velocity, and that do not lie at the tangent point, we
derive a KDAR for \input kda_resolved ($\input kda_percent\%$).  This
percentage is lower than that of the HRDS \citep[67\% if tangent-point
  sources are excluded;][]{anderson12c}.  

We give the results of our analysis of the discrete \hii\ regions in
Table~\ref{tab:distances}, which lists the source LSR velocity, the
near, far, and tangent point kinematic distances, the KDAR, the
quality factor for the KDAR, the Galactocentric radius, the
heliocentric distance, the height above the plane, and the criteria
used to determine the \hii\ region velocity.

\section{Summary}
We identify the discrete \hii\ region velocity for \input resolved
($\input percentresolved\%$) \hii\ regions whose radio recombination
line (RRL) spectra show multiple velocity components.  We do this by
using new Green Bank Telescope (GBT) observations, analyzing the derived
electron temperature for each velocity component, and searching for
the molecular emission or carbon recombination lines associated with
one RRL component.  

Our sample contains all multiple-velocity \hii\ regions in the zone
$18\degree < \ell < 65\degree$.  The multiple-velocity regions are
concentrated in the zone $22\degree < \ell < 32\degree$, spatially
coincident with some of the largest regions of massive star formation
in the Milky Way.  This suggests that the additional velocity
components along these sight lines may be due to leaked photons from
large massive star formation complexes.  We thus interpret these
additional components as stemming from diffuse gas ionized by photons
from these energetic regions.

Prior to our analyses, it was not possible to derive distances to
these regions, or to derive their physical properties.  We derive
kinematic distances for \input ndistance \hii\ regions, \input
ntangent at the tangent point distance and \input kda_resolved for
which we resolved the kinematic distance ambiguity.  This
  population of \hii\ regions can now be used for large-scale Galactic
  studies involving \hii\ regions, including studies of Galactic
  structure (Bania et al., 2015, in prep.), metallicity (Balser et
  al., 2015, submitted; Wenger et al., 2015, in prep.), and the star
  formation rate (Anderson et al., 2015, in prep.)

Finally, our analysis leads naturally to the creation of a large
database of diffuse RRL line parameters.  Unlike most previous low
angular resolution radio investigations of the diffuse RRL emission in
our Galaxy, these data sample only the diffuse RRL component, and are
uncontaminated by discrete \hii\ regions.  We will investigate
this rich database in a future paper.

\begin{acknowledgments}
\nraoblurb\ We thank the staff at Green Bank for their friendship and
hospitality during work on this project.  We also thank the anonymous
referee, whose comments improved the clarity of this manuscript.  This
project made use of the WISE Catalog of Galactic HII Regions, made
possible by NASA ADAP grant NNX12AI59G to LDA.\\

{\it Facility: Green Bank Telescope}
\end{acknowledgments}

\bibliographystyle{apj} 
\bibliography{../../ref.bib} 

\clearpage
\begin{figure}
\includegraphics[width=6.5in]{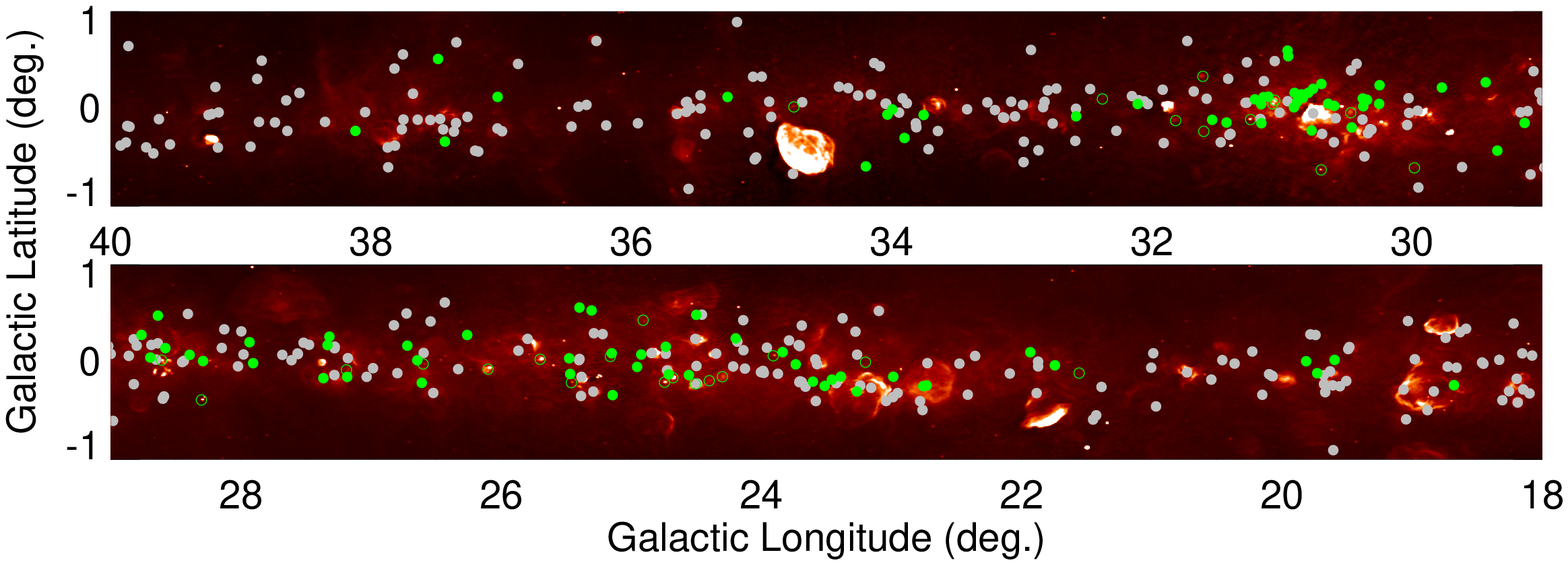}
\caption{Locations of the single- and multiple-velocity \hii\ regions.  The
  background image is a mosaic of VGPS 21\,cm continuum data
  \citep{stil06}.  Multiple-velocity \hii\ regions are shown in green,
  with HRDS sources filled and those from the literature open.  Single
  velocity \hii\ regions are shown in light gray.  The
  multiple-velocity \hii\ regions are not evenly distributed and are
  rare below $\ell \simeq 22\degree$ and above $\ell \simeq
  32\degree$.  One multiple-velocity \hii\ region near $\ell =
    50\degree$ is not shown.} \label{fig:multvel_pos}
\end{figure}

\clearpage
\begin{figure}
\begin{centering}
\includegraphics[height=1.6in]{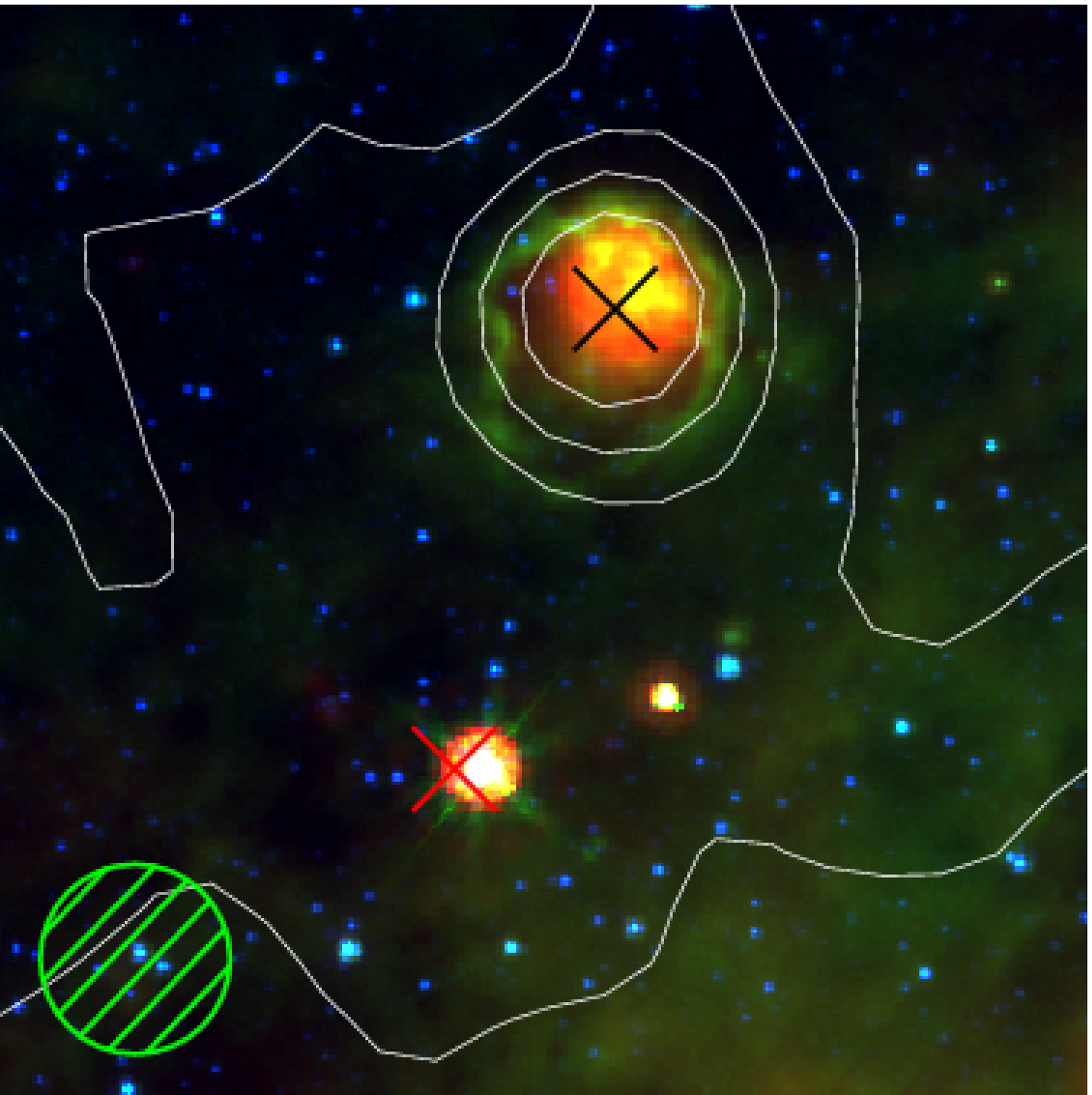}
\includegraphics[height=1.6in]{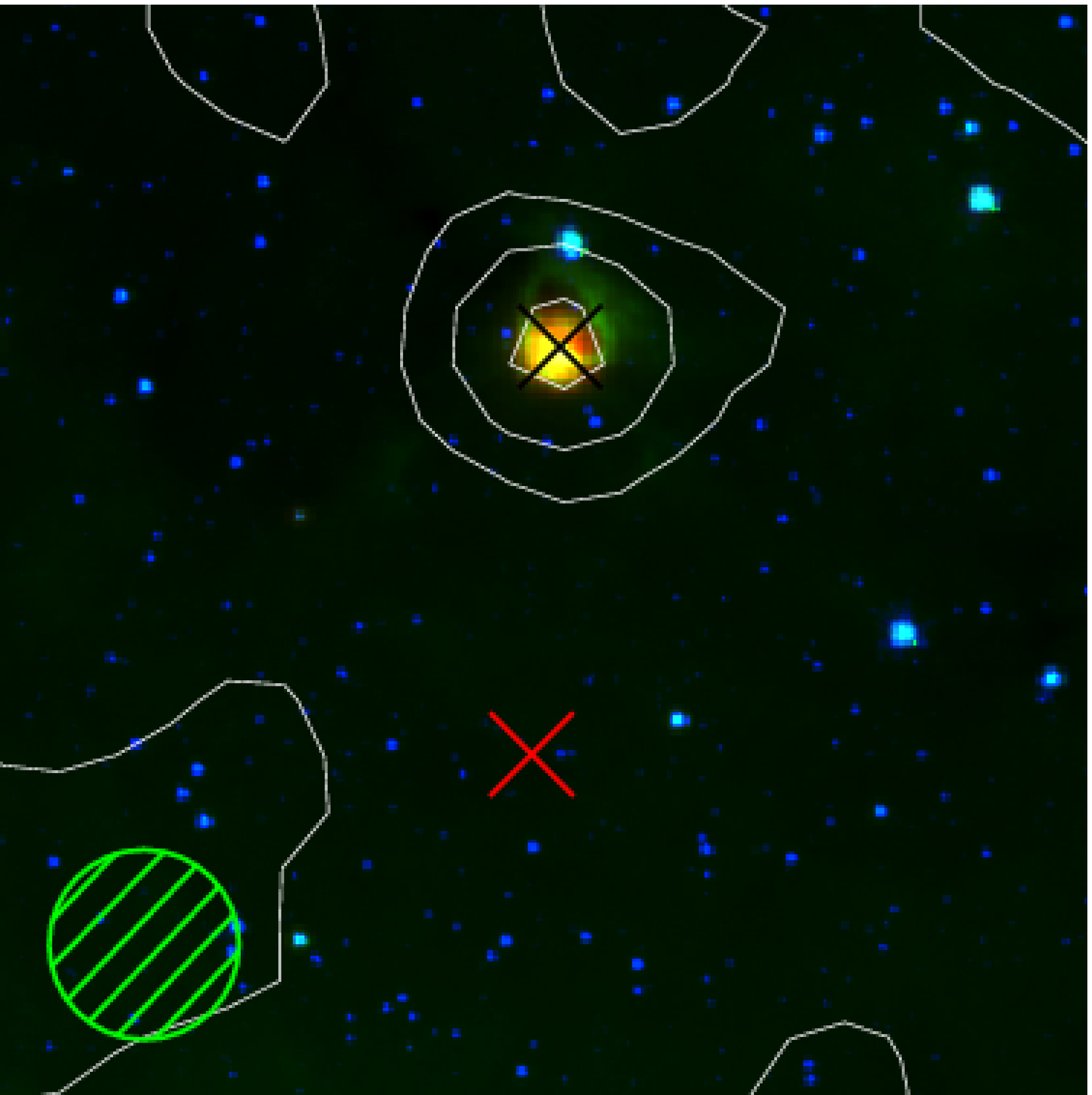}
\caption{Example ``off-target'' locations for G030.249+0.243 (left)
  and G025.305+0.532 (right).  The images show emission from {\it
    Spitzer} GLIMPSE 3.6\,\micron\ (blue), GLIMPSE
  8.0\,\micron\ (green), and MIPSGAL 24\,\micron\ (red), and are $7.5\arcmin$ square.  VGPS 21\,cm
  continuum contours are over-plotted in white and the GBT beam at
  X-band is in the lower left of each image.  The black crosses show
  the original HRDS positions and the red crosses show the off-target
  positions.  (The infrared source nearly coincident with the red
  cross for G30.249+0.243 has no associated radio continuum and
  therefore did not affect our observations).}\label{fig:offtarget}
\end{centering}
\end{figure}

\clearpage
\begin{figure}
\begin{centering}
\includegraphics[width=3in]{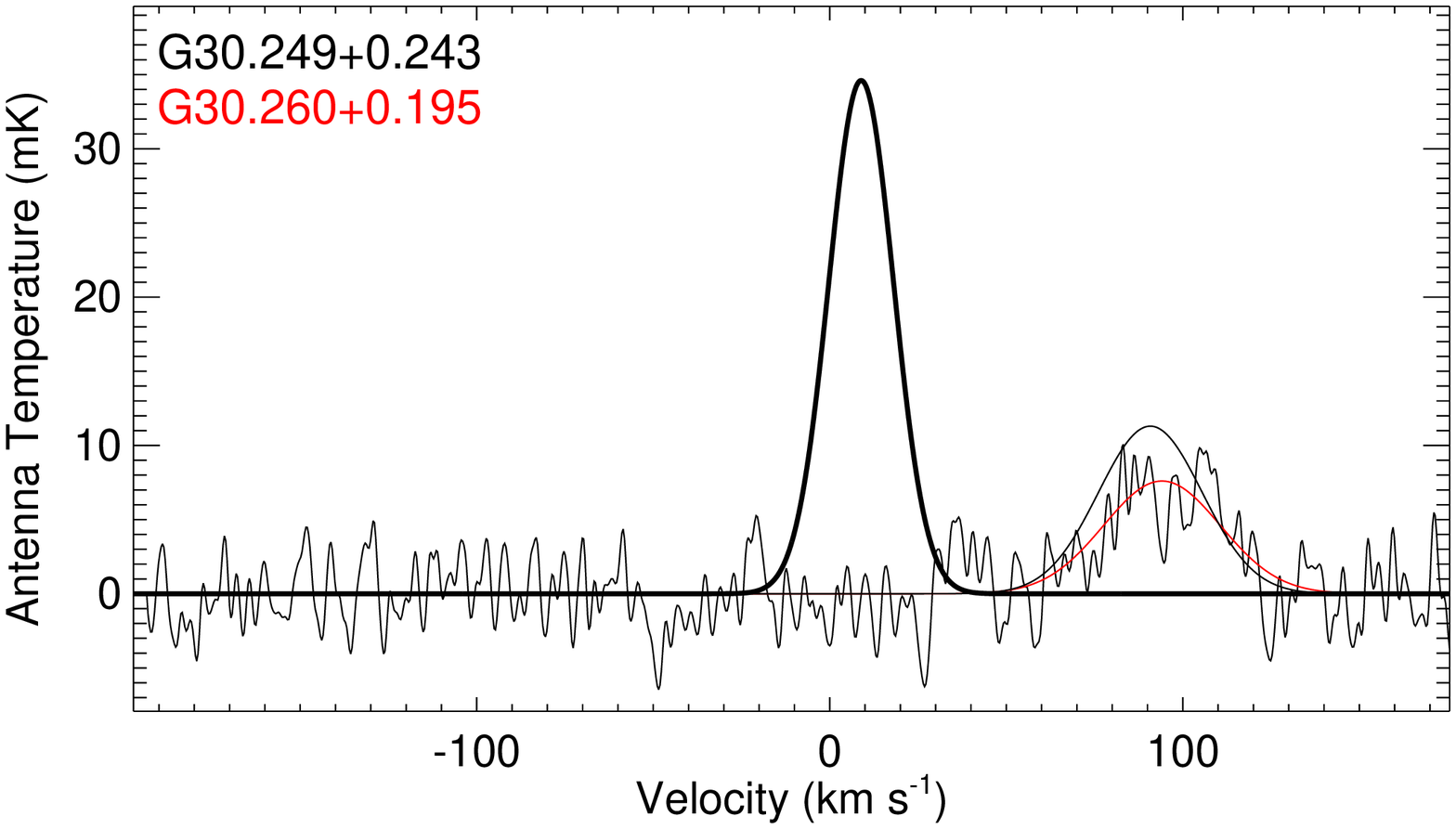}
\includegraphics[width=3in]{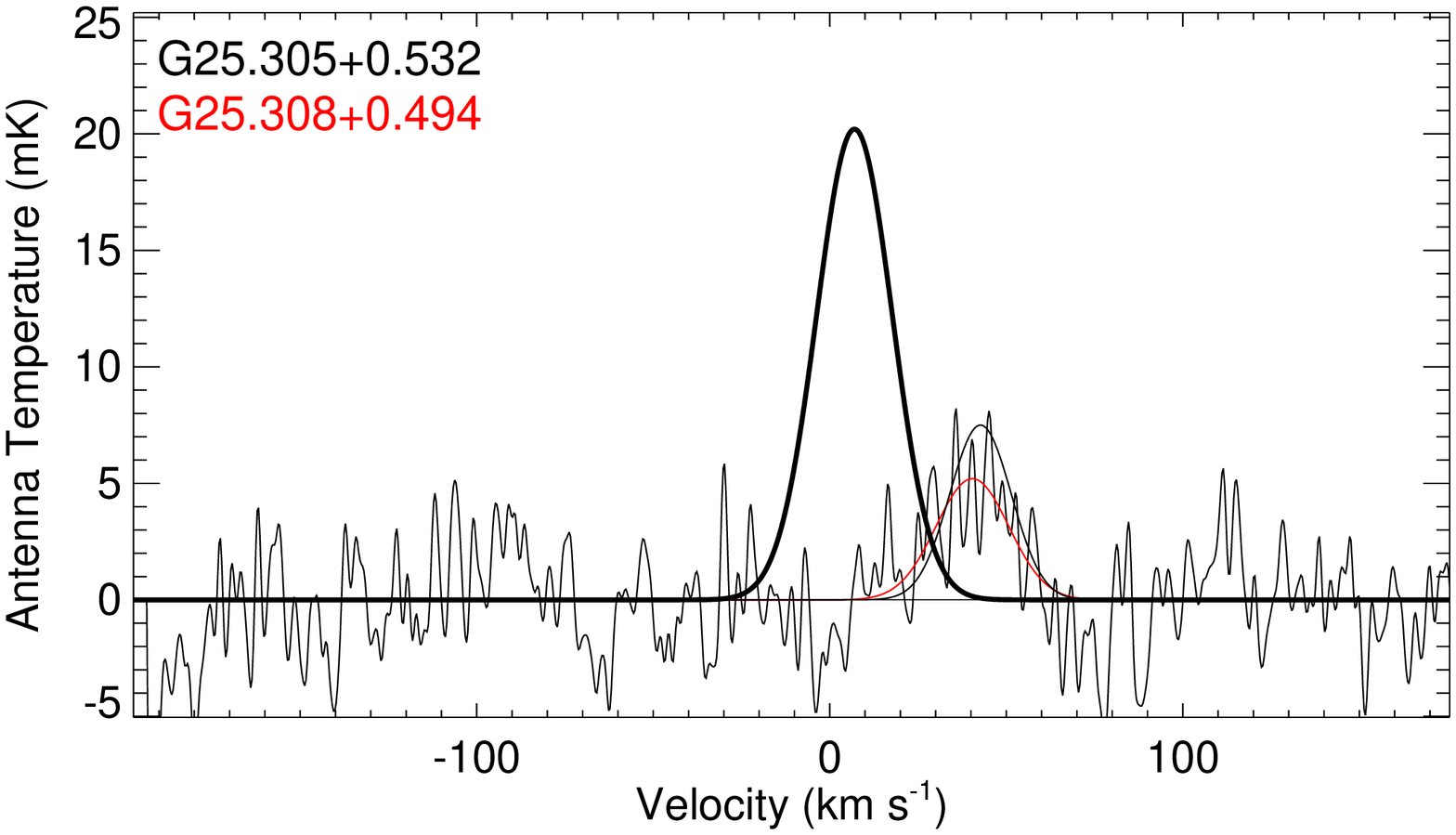}
\caption{RRL component intensity comparison for on- and
    off-target spectra.  Shown is the intensity comparison for the
    Figure~\ref{fig:offtarget} \hii\ region targets.  The black curves
    are Gaussian fits to the line components found in the on-target
    positions of the original HRDS (from black crosses in
    Figure~\ref{fig:offtarget}).  The spectra are those of the
    off-target positions (red crosses in Figure~\ref{fig:offtarget})
    and the red curves are Gaussian fits to these data.  In both cases, the
    line components near 0\,\kms\ are not detected at the off-target
    locations.  These components are therefore from the discrete
    \hii\ regions.} \label{fig:spectra}
\end{centering}
\end{figure}

\clearpage
\begin{figure}
\begin{centering}
\includegraphics[width=4.5in]{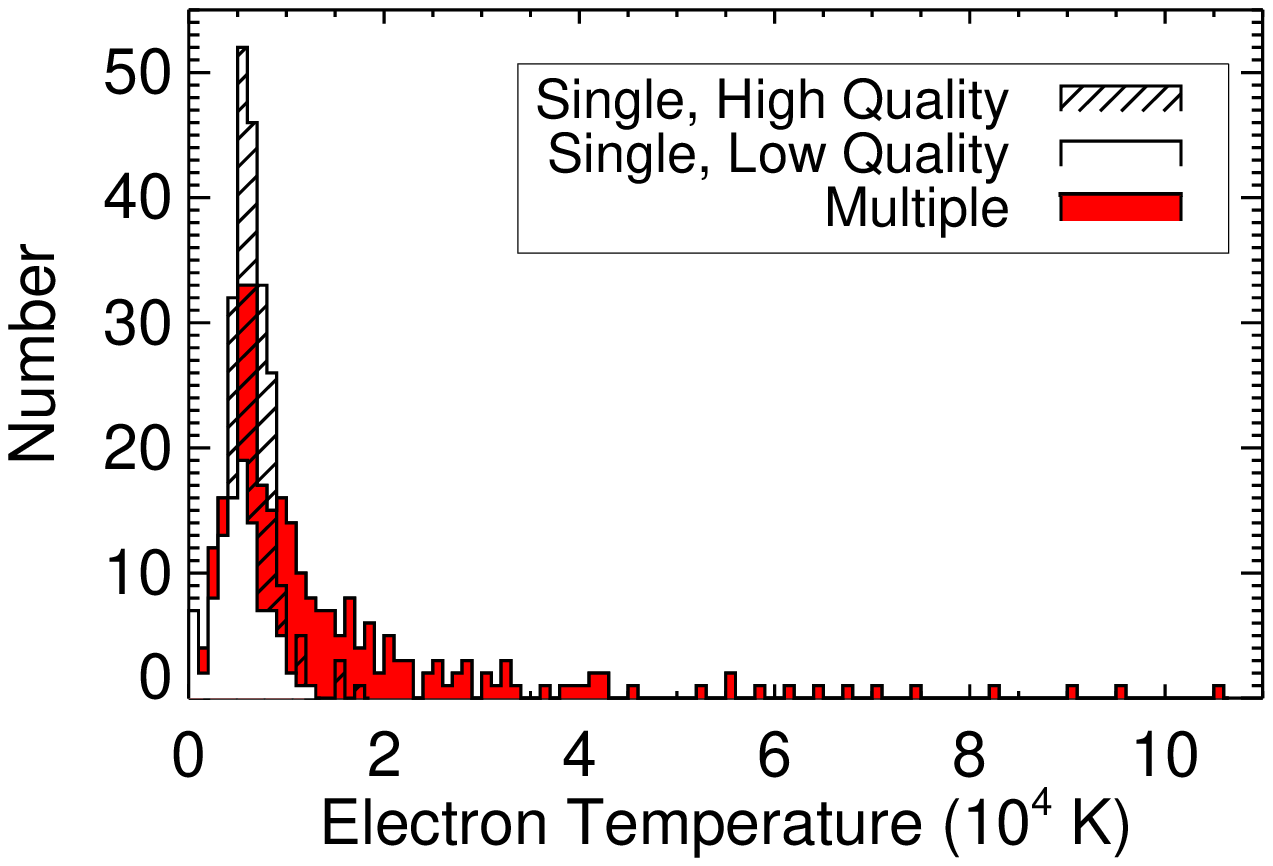}
\includegraphics[width=4.5in]{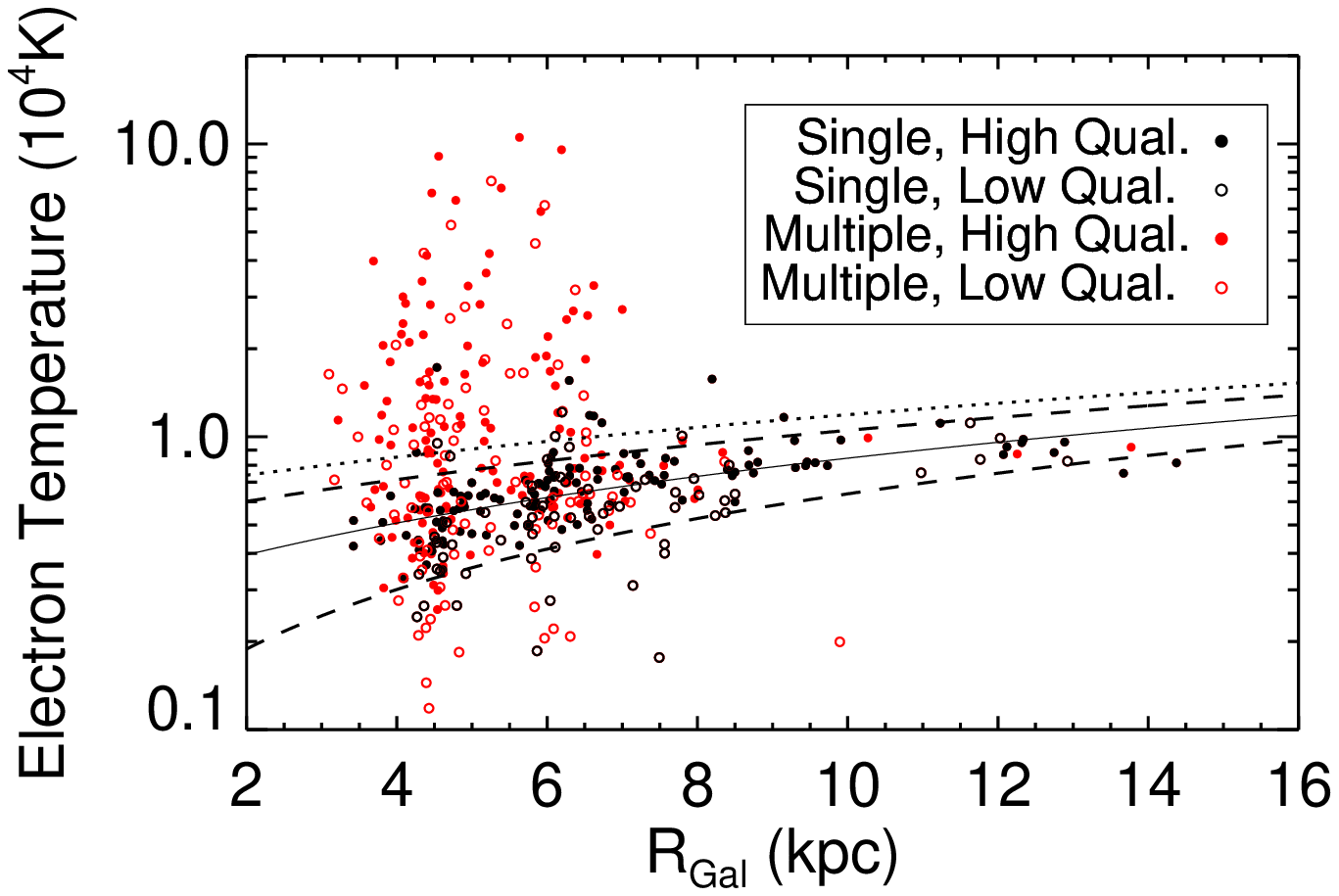}
\caption{Electron temperatures for single- and multiple-velocity
  \hii\ regions.  {\it Top:} the electron temperature distributions
  for all HRDS \hii\ regions.  The single- and multiple-velocity
  sources have different distributions.  No single-velocity HRDS
  source has an electron temperature greater than $2\times 10^4$\,K.
  The vast majority of HRDS sources with electron temperatures greater
  than $10^4$\,K have multiple velocity components.  {\it Bottom:} the
  electron temperature distribution for HRDS sources within $18\degree
  < \ell < 65\degree$, as a function of Galactocentric radius.  The
  zones outside of the dashed lines are unreasonable for high-quality
  sources, as is the zone above the dotted line for low-quality
  sources (see text).  The line components in these zones are unlikely
  to be from discrete \hii\ regions. \label{fig:te}}
\end{centering}
\end{figure}

\clearpage
\begin{figure}
  \includegraphics[width=4.5in]{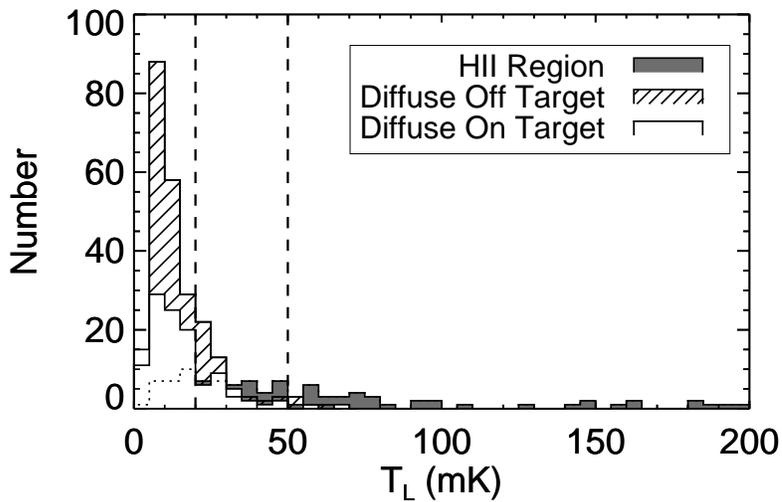}
  \caption{Peak RRL intensity distribution for diffuse gas and
    discrete \hii\ regions.  The on- and off-target diffuse component
    intensities share a similar distribution, whereas the
    \hii\ regions are significantly brighter.  Vertical dashed lines
    show the 20\,mK and 50\,mK criteria we use when determining the
    source velocities.  The dotted line is the low-intensity portion
    of the dark grey discrete \hii\ region
    distribution.\label{fig:height}}
\end{figure}

\clearpage
\begin{deluxetable}{lccccccccc}
\tabletypesize{\scriptsize}
\tablecaption{On-Target Hydrogen RRL Parameters of Literature HII Regions}
\tablewidth{0pt}
\tablehead{
\colhead{Source} &
\colhead{$\ell$} &
\colhead{$b$} &
\colhead{$T_L$} &
\colhead{$\sigma_{T\!L}$} &
\colhead{$V_{lsr}$} &
\colhead{$\sigma_{Vlsr}$} &
\colhead{$\Delta V$} &
\colhead{$\sigma_{\Delta V}$} &
\colhead{rms}
\\
\colhead{} &
\colhead{deg.} &
\colhead{deg.} &
\colhead{\mK} &
\colhead{\mK} &
\colhead{\kms} &
\colhead{\kms} &
\colhead{\kms} &
\colhead{\kms} &
\colhead{\mK}
}
\startdata
\input gsource.tab
\enddata
\label{tab:literature}
\end{deluxetable}


\clearpage
\begin{deluxetable}{lcccccccccl}
\rotate
\tabletypesize{\scriptsize}
\setlength{\tabcolsep}{1pt}
\tablecaption{Off-Target Hydrogen RRL Parameters for the Full Sample}
\tablewidth{0pt}
\tablehead{
\colhead{$\ell$} &
\colhead{$b$} &
\colhead{Separation$^{\rm a}$} &
\colhead{$T_L$} &
\colhead{$\sigma_{TL}$} &
\colhead{$V_{lsr}$} &
\colhead{$\sigma_{vlsr}$} &
\colhead{$\Delta V$} &
\colhead{$\sigma_{\Delta V}$} &
\colhead{rms} &
\colhead{HII Region} 
\\
\colhead{deg.} &
\colhead{deg.} &
\colhead{arcmin.} &
\colhead{\mK} &
\colhead{\mK} &
\colhead{\kms} &
\colhead{\kms} &
\colhead{\kms} &
\colhead{\kms} &
\colhead{\mK} &
\colhead{}
}
\startdata
\input diffuse.tab
\enddata

\tablenotetext{a}{From the nominal centroid position of the nearest multiple-velocity HII region.}

\label{tab:offtarget}
\end{deluxetable}

\newpage
\begin{deluxetable}{lccl}
\tabletypesize{\scriptsize}
\tablecaption{Efficacy of Criteria}
\tablewidth{0pt}
\tablehead{
\colhead{Criterion} &
\colhead{Number} &
\colhead{Percentage} &
\colhead{Comments}
\\
\colhead{} &
\colhead{of Sources} &
\colhead{of Sources} &
\colhead{}
}
\startdata
\input reasons.tab
\enddata


\label{tab:criteria}
\end{deluxetable}

\newpage
\begin{deluxetable}{lcccccccrrr}
\tabletypesize{\scriptsize}
\tablecaption{HII Region Velocities and Distances}
\tablewidth{0pt}
\tablehead{
\colhead{Source} &
\colhead{$V_{\rm lsr}$} &
\colhead{$D_N$} &
\colhead{$D_F$} &
\colhead{$D_{TP}$} &
\colhead{KDAR} &
\colhead{QF} &
\colhead{$R_{gal}$} &
\colhead{$D_\sun$} &
\colhead{$z$} &
\colhead{criterion$^{\rm a}$} 
\\
\colhead{} &
\colhead{\kms} &
\colhead{\kpc} &
\colhead{\kpc} &
\colhead{\kpc} &
\colhead{} &
\colhead{} &
\colhead{\kpc} &
\colhead{\kpc} &
\colhead{\pc} &
\colhead{}
}
\startdata
\input distances_lda.tab
\enddata

\tablenotetext{a}{See text.}

\label{tab:distances}
\end{deluxetable}

\end{document}

%% file: n
117

%% file: resolved
103

%% file: percentresolved
88

%% file: kda_resolved
47

%% file: nliterature
28

%% file: nhrds
89

%% file: nliteraturetotal
53

%% file: reason1
5

%% file: noffpointing
127

%% file: noff
225

%% file: ntotal
474

%% file: nhrdstotal
196

%% file: singletestats
$6700\pm2100$

%% file: singlecomplextestats
$5300\pm2600$

%% file: multipletestats
$14000\pm16000$

%% file: singletepercentage
          91

%% file: singletecomplexpercentage
          95

%% file: kda_possible
86

%% file: kda_percent
55

%% file: ndistance
62

%% file: ntangent
15

%% file: gsource.tab
G021.558$-$0.112 & 21.558 & $-$0.112 & 24.9 & 0.3 & 115.1 & 0.2 & 23.0 & 0.5 & 2.2 \\ 
G023.200+0.000 & 23.200 & 0.000 & 13.4 & 0.4 & 89.5 & 0.5 & 33.0 & 1.5 & 2.3 \\ 
G023.909+0.066a & 23.909 & 0.066 & 49.0 & 0.3 & 38.4 & 0.1 & 24.5 & 0.3 & 2.2 \\ 
G023.909+0.066b & 23.909 & 0.066 & 33.0 & 0.6 & 107.9 & 0.2 & 26.6 & 0.5 & 2.2 \\ 
G024.300$-$0.149a & 24.300 & $-$0.149 & 70.0 & 0.5 & 56.8 & 0.1 & 17.5 & 0.1 & 2.4 \\ 
G024.300$-$0.149b & 24.300 & $-$0.149 & 30.1 & 0.5 & 95.0 & 0.1 & 19.1 & 0.3 & 2.4 \\ 
G024.400$-$0.190a & 24.400 & $-$0.190 & 45.1 & 0.6 & 101.0 & 0.2 & 21.8 & 0.4 & 2.4 \\ 
G024.400$-$0.190b & 24.400 & $-$0.190 & 11.9 & 0.6 & 54.7 & 0.7 & 27.9 & 1.7 & 2.4 \\ 
G024.507$-$0.222a & 24.507 & $-$0.222 & 232.4 & 1.6 & 96.3 & 0.1 & 23.7 & 0.2 & 8.5 \\ 
G024.507$-$0.222b & 24.507 & $-$0.222 & 25.3 & 1.8 & 42.2 & 0.6 & 17.2 & 1.4 & 8.5 \\ 
G024.680$-$0.160a & 24.680 & $-$0.160 & 421.1 & 0.5 & 110.3 & 0.1 & 23.4 & 0.1 & 2.1 \\ 
G024.680$-$0.160b & 24.680 & $-$0.160 & 44.9 & 0.6 & 43.6 & 0.1 & 13.1 & 0.2 & 2.1 \\ 
G024.744$-$0.206a & 24.744 & $-$0.206 & 162.7 & 0.8 & 82.6 & 0.1 & 19.5 & 0.1 & 2.9 \\ 
G024.744$-$0.206b & 24.744 & $-$0.206 & 24.6 & 1.0 & 44.0 & 0.3 & 13.5 & 0.6 & 2.9 \\ 
G024.744$-$0.206c & 24.744 & $-$0.206 & 17.1 & 0.4 & 122.6 & 0.2 & 17.8 & 0.6 & 2.9 \\ 
G024.909+0.432a & 24.909 & 0.432 & 28.8 & 0.7 & 108.6 & 0.3 & 18.2 & 0.6 & 2.6 \\ 
G024.909+0.432b & 24.909 & 0.432 & 13.9 & 0.5 & 135.1 & 0.8 & 27.4 & 1.9 & 2.6 \\ 
G025.160+0.059a & 25.160 & 0.059 & 66.3 & 0.4 & 45.5 & 0.1 & 25.1 & 0.2 & 2.4 \\ 
G025.160+0.059b & 25.160 & 0.059 & 9.2 & 0.4 & 107.7 & 0.5 & 24.8 & 1.5 & 2.4 \\ 
G025.460$-$0.210a & 25.460 & $-$0.210 & 56.6 & 0.4 & 117.3 & 0.1 & 30.5 & 0.2 & 2.2 \\ 
G025.460$-$0.210b & 25.460 & $-$0.210 & 15.4 & 0.4 & 62.5 & 0.4 & 31.6 & 0.9 & 2.2 \\ 
G025.700+0.030a & 25.700 & 0.030 & 195.4 & 0.9 & 52.7 & 0.1 & 21.3 & 0.1 & 2.9 \\ 
G025.700+0.030b & 25.700 & 0.030 & 49.6 & 1.2 & 124.2 & 0.2 & 13.2 & 0.4 & 2.9 \\ 
G025.700+0.030c & 25.700 & 0.030 & 19.8 & 0.4 & 83.4 & 0.4 & 20.7 & 1.6 & 2.9 \\ 
G026.100$-$0.071a & 26.100 & $-$0.071 & 77.2 & 0.6 & 30.7 & 0.1 & 27.4 & 0.3 & 2.3 \\ 
G026.100$-$0.071b & 26.100 & $-$0.071 & 25.6 & 0.6 & 103.4 & 0.3 & 30.0 & 0.8 & 2.3 \\ 
G026.599$-$0.019a & 26.599 & $-$0.019 & 33.3 & 0.5 & 19.4 & 0.2 & 25.7 & 0.4 & 2.1 \\ 
G026.599$-$0.019b & 26.599 & $-$0.019 & 10.9 & 0.4 & 102.3 & 0.6 & 33.5 & 1.4 & 2.1 \\ 
G027.189$-$0.079a & 27.189 & $-$0.079 & 20.0 & 0.3 & 27.8 & 0.2 & 22.8 & 0.5 & 1.9 \\ 
G027.189$-$0.079b & 27.189 & $-$0.079 & 10.6 & 0.3 & 86.9 & 0.4 & 34.7 & 1.1 & 1.9 \\ 
G028.303$-$0.388 & 28.303 & $-$0.388 & 148.7 & 0.5 & 76.9 & 0.1 & 21.2 & 0.1 & 2.9 \\ 
G028.609+0.021a & 28.609 & 0.021 & 161.2 & 0.6 & 96.6 & 0.1 & 20.5 & 0.1 & 3.3 \\ 
G028.609+0.021b & 28.609 & 0.021 & 10.0 & 0.6 & 40.3 & 0.5 & 15.4 & 1.1 & 3.3 \\ 
G029.981$-$0.607 & 29.981 & $-$0.607 & 45.1 & 0.4 & 90.9 & 0.1 & 19.6 & 0.2 & 1.9 \\ 
G030.470$-$0.040a & 30.470 & $-$0.040 & 62.9 & 0.7 & 42.7 & 0.1 & 22.1 & 0.3 & 2.6 \\ 
G030.470$-$0.040b & 30.470 & $-$0.040 & 33.8 & 0.5 & 101.4 & 0.3 & 32.3 & 0.6 & 2.6 \\ 
G030.698$-$0.628a & 30.698 & $-$0.628 & 19.0 & 0.4 & 87.8 & 0.3 & 16.5 & 0.9 & 2.0 \\ 
G030.698$-$0.628b & 30.698 & $-$0.628 & 8.6 & 0.4 & 110.2 & 0.8 & 20.1 & 2.4 & 2.0 \\ 
G031.054+0.088a & 31.054 & 0.088 & 95.3 & 0.7 & 20.1 & 0.1 & 28.2 & 0.2 & 2.9 \\ 
G031.054+0.088b & 31.054 & 0.088 & 33.0 & 0.6 & 97.8 & 0.4 & 36.1 & 1.0 & 2.9 \\ 
G031.070+0.050a & 31.070 & 0.050 & 109.5 & 0.6 & 35.4 & 0.1 & 29.0 & 0.2 & 2.9 \\ 
G031.070+0.050b & 31.070 & 0.050 & 39.8 & 0.5 & 98.6 & 0.2 & 34.6 & 0.5 & 2.9 \\ 
G031.240$-$0.110a & 31.240 & $-$0.110 & 155.6 & 0.9 & 26.5 & 0.1 & 29.5 & 0.2 & 3.0 \\ 
G031.240$-$0.110b & 31.240 & $-$0.110 & 18.1 & 0.5 & 90.3 & 0.5 & 29.5 & 1.5 & 3.0 \\ 
G031.600$-$0.232a & 31.600 & $-$0.232 & 35.6 & 0.4 & 40.8 & 0.1 & 19.6 & 0.3 & 2.5 \\ 
G031.600$-$0.232b & 31.600 & $-$0.232 & 12.6 & 0.4 & 100.9 & 0.3 & 20.0 & 0.7 & 2.5 \\ 
G031.607+0.334a & 31.607 & 0.334 & 44.9 & 0.4 & 24.4 & 0.1 & 24.3 & 0.2 & 2.6 \\ 
G031.607+0.334b & 31.607 & 0.334 & 6.1 & 0.4 & 96.4 & 0.9 & 28.8 & 2.2 & 2.6 \\ 
G031.816$-$0.121a & 31.816 & $-$0.121 & 38.8 & 0.5 & 36.3 & 0.1 & 20.8 & 0.3 & 2.6 \\ 
G031.816$-$0.121b & 31.816 & $-$0.121 & 11.6 & 0.5 & 104.3 & 0.4 & 21.6 & 1.1 & 2.6 \\ 
G032.379+0.102 & 32.379 & 0.102 & 15.2 & 0.3 & 47.3 & 0.3 & 25.8 & 0.7 & 1.9 \\ 
G034.750+0.020a & 34.750 & 0.020 & 36.0 & 0.5 & 86.1 & 0.1 & 20.2 & 0.3 & 2.7 \\ 
G034.750+0.020b & 34.750 & 0.020 & 25.4 & 0.5 & 52.0 & 0.2 & 21.0 & 0.5 & 2.7 \\ 

%% file: diffuse.tab
18.695 & $-$0.274 & 2.5&15.7 & 0.6 & 64.8 & 1.0 & 23.4 & 2.0 & 3.5 & G018.677$-$00.236 \\ 
 &  & &11.7 & 0.8 & 40.5 & 1.1 & 19.4 & 2.2 & 3.5 &  \\ 
19.718 & 0.178 & 11.8&8.3 & 0.4 & 122.4 & 0.3 & 15.1 & 0.9 & 2.6 & G019.594+00.024 \\ 
19.722 & $-$0.078 & 2.1&13.5 & 0.4 & 116.3 & 0.4 & 22.5 & 1.1 & 2.4 & G019.728$-$00.113 \\ 
 &  & &11.1 & 0.5 & 56.9 & 0.5 & 22.5 & 1.1 & 2.4 &  \\ 
19.822 & $-$0.030 & 2.5&11.9 & 0.5 & 120.0 & 0.3 & 15.5 & 0.8 & 2.3 & G019.813+00.010 \\ 
 &  & &6.6 & 0.3 & 61.9 & 0.5 & 21.3 & 1.3 & 2.3 &  \\ 
21.916 & 0.064 & 2.5&5.1 & 0.3 & 60.5 & 1.1 & 31.0 & 3.8 & 2.3 & G021.933+00.103 \\ 
21.985 & 0.137 & 3.8&4.7 & 0.3 & 57.2 & 1.5 & 31.9 & 4.9 & 2.2 & G021.933+00.103 \\ 
22.700 & $-$0.269 & 2.5&13.9 & 0.5 & 71.2 & 0.4 & 20.2 & 1.0 & 2.5 & G022.730$-$00.239, G022.755$-$00.246 \\ 
 &  & &5.9 & 0.6 & 111.2 & 0.7 & 14.7 & 1.8 & 2.5 &  \\ 
22.738 & $-$0.189 & 3.0&14.7 & 0.4 & 72.5 & 0.3 & 27.4 & 0.8 & 2.7 & G022.730$-$00.239, G022.755$-$00.246 \\ 
22.811 & $-$0.202 & 4.3&21.2 & 0.3 & 71.3 & 0.2 & 23.7 & 0.5 & 2.2 & G022.730$-$00.239, G022.755$-$00.246 \\ 
 &  & &14.6 & 0.4 & 112.9 & 0.2 & 13.6 & 0.6 & 2.2 &  \\ 
 &  & &14.4 & 0.5 & 95.1 & 0.2 & 10.8 & 0.6 & 2.2 &  \\ 
22.930 & $-$0.149 & 3.3&10.6 & 0.2 & 79.5 & 0.6 & 44.1 & 1.3 & 1.8 & G022.986$-$00.149 \\ 
 &  & &8.5 & 0.5 & 101.5 & 0.2 & 8.8 & 0.7 & 1.8 &  \\ 
23.004 & $-$0.194 & 2.9&21.4 & 0.4 & 71.6 & 0.3 & 19.6 & 0.8 & 2.1 & G022.986$-$00.149 \\ 
 &  & &11.9 & 0.4 & 94.4 & 0.6 & 19.5 & 1.3 & 2.1 &  \\ 
23.037 & $-$0.117 & 3.6&9.1 & 0.2 & 81.2 & 0.7 & 44.9 & 1.9 & 2.1 & G022.986$-$00.149 \\ 
23.225 & $-$0.344 & 3.5&17.3 & 0.3 & 78.6 & 0.4 & 41.5 & 1.0 & 1.9 & G023.265$-$00.301 \\ 
23.372 & $-$0.110 & 2.5&19.6 & 0.3 & 90.7 & 0.2 & 20.6 & 0.4 & 2.1 & G023.389$-$00.148 \\ 
 &  & &9.8 & 0.4 & 55.1 & 0.3 & 15.0 & 0.7 & 2.1 &  \\ 
23.481 & $-$0.216 & 2.6&19.6 & 0.4 & 96.0 & 0.2 & 19.0 & 0.5 & 2.2 & G023.458$-$00.179 \\ 
 &  & &14.6 & 0.4 & 57.1 & 0.3 & 18.4 & 0.7 & 2.2 &  \\ 
23.588 & $-$0.239 & 2.5&9.3 & 0.3 & 95.4 & 0.9 & 36.0 & 2.4 & 2.4 & G023.604$-$00.200 \\ 
 &  & &5.8 & 0.4 & 55.5 & 1.2 & 25.1 & 3.4 & 2.4 &  \\ 
23.720 & $-$0.055 & 2.2&13.1 & 0.6 & 106.1 & 0.5 & 21.9 & 1.2 & 3.3 & G023.736$-$00.022 \\ 
23.845 & 0.061 & 2.6&13.7 & 0.3 & 106.3 & 0.6 & 41.8 & 1.8 & 1.9 & G023.836+00.104 \\ 
23.860 & 0.037 & 3.4&20.5 & 0.5 & 117.2 & 0.2 & 12.6 & 0.4 & 2.6 & G023.909+00.066 \\ 
 &  & &11.4 & 0.3 & 98.5 & 0.5 & 23.9 & 1.7 & 2.6 &  \\ 
24.158 & 0.242 & 2.3&22.6 & 0.8 & 111.5 & 0.3 & 17.1 & 0.7 & 2.3 & G024.197+00.245 \\ 
24.346 & $-$0.139 & 2.9&22.8 & 0.4 & 94.0 & 0.1 & 19.0 & 0.3 & 2.0 & G024.300$-$00.149 \\ 
 &  & &16.1 & 0.4 & 54.4 & 0.2 & 18.0 & 0.5 & 2.0 &  \\ 
24.369 & $-$0.225 & 2.8&24.8 & 0.4 & 96.4 & 0.2 & 18.9 & 0.4 & 1.8 & G024.400$-$00.190 \\ 
 &  & &7.0 & 0.3 & 54.4 & 0.7 & 29.2 & 1.8 & 1.8 &  \\ 
24.425 & $-$0.177 & 1.7&28.0 & 0.4 & 100.3 & 0.2 & 24.8 & 0.4 & 2.1 & G024.400$-$00.190 \\ 
 &  & &6.4 & 0.4 & 54.4 & 0.7 & 22.0 & 1.7 & 2.1 &  \\ 
24.476 & $-$0.198 & 2.4&28.1 & 0.4 & 94.7 & 0.2 & 20.4 & 0.4 & 2.2 & G024.507$-$00.222 \\ 
 &  & &9.6 & 0.4 & 46.6 & 0.5 & 26.9 & 1.2 & 2.2 &  \\ 
 &  & &8.7 & 0.4 & 123.8 & 0.5 & 17.8 & 1.3 & 2.2 &  \\ 
24.498 & $-$0.182 & 2.5&45.2 & 2.5 & 89.9 & 0.2 & 7.8 & 0.5 & 7.6 & G024.507$-$00.222 \\ 
 &  & &20.7 & 2.2 & 52.9 & 0.5 & 10.0 & 1.2 & 7.6 &  \\ 
24.512 & 0.538 & 3.2&11.6 & 0.5 & 34.2 & 0.3 & 11.9 & 0.7 & 2.0 & G024.500+00.487 \\ 
 &  & &8.8 & 0.4 & 100.5 & 0.4 & 17.3 & 1.0 & 2.0 &  \\ 
24.518 & $-$0.152 & 2.6&23.8 & 0.4 & 92.3 & 0.2 & 27.1 & 0.6 & 2.1 & G024.557$-$00.132 \\ 
 &  & &10.7 & 0.4 & 52.3 & 0.5 & 27.9 & 1.5 & 2.1 &  \\ 
24.634 & $-$0.183 & 3.1&51.9 & 0.8 & 42.5 & 0.1 & 13.1 & 0.2 & 2.6 & G024.680$-$00.160 \\ 
 &  & &13.1 & 0.4 & 109.9 & 1.1 & 53.4 & 2.8 & 2.6 &  \\ 
24.637 & $-$0.109 & 4.7&59.9 & 1.5 & 104.5 & 0.3 & 23.4 & 0.7 & 2.7 & G024.713$-$00.125 \\ 
 &  & &21.0 & 0.9 & 42.4 & 0.2 & 10.2 & 0.6 & 2.7 &  \\ 
24.638 & $-$0.183 & 2.9&51.8 & 0.8 & 42.5 & 0.1 & 13.1 & 0.2 & 2.5 & G024.680$-$00.160 \\ 
 &  & &12.8 & 0.4 & 109.7 & 1.1 & 52.2 & 2.8 & 2.5 &  \\ 
24.694 & 0.144 & 2.6&23.1 & 0.4 & 109.6 & 0.2 & 23.3 & 0.6 & 2.1 & G024.735+00.159 \\ 
 &  & &12.7 & 0.4 & 81.4 & 0.2 & 14.1 & 0.6 & 2.1 &  \\ 
24.784 & $-$0.178 & 2.9&22.9 & 0.5 & 40.9 & 0.1 & 11.4 & 0.3 & 2.5 & G024.744$-$00.206 \\ 
 &  & &12.7 & 0.3 & 113.8 & 0.4 & 29.0 & 1.0 & 2.5 &  \\ 
24.799 & $-$0.151 & 5.4&15.4 & 0.4 & 110.4 & 0.4 & 30.8 & 1.0 & 2.1 & G024.713$-$00.125 \\ 
 &  & &15.1 & 0.6 & 40.9 & 0.3 & 16.5 & 0.8 & 2.1 &  \\ 
24.902 & $-$0.065 & 3.3&13.4 & 0.4 & 99.0 & 0.5 & 28.8 & 1.1 & 2.0 & G024.955$-$00.048 \\ 
 &  & &8.0 & 0.4 & 50.9 & 0.5 & 16.9 & 1.4 & 2.0 &  \\ 
24.906 & 0.039 & 2.6&14.7 & 0.4 & 109.7 & 0.3 & 23.6 & 0.8 & 2.3 & G024.923+00.079 \\ 
24.965 & 0.391 & 4.2&11.9 & 0.4 & 111.1 & 0.3 & 18.2 & 0.7 & 1.9 & G024.909+00.432 \\ 
25.128 & 0.121 & 2.2&15.8 & 0.5 & 95.2 & 0.4 & 25.7 & 1.2 & 2.3 & G025.150+00.092 \\ 
25.155 & $-$0.372 & 2.2&6.2 & 0.5 & 123.9 & 0.6 & 15.0 & 1.4 & 2.3 & G025.143$-$00.338 \\ 
25.205 & 0.045 & 2.8&9.6 & 0.3 & 50.9 & 0.6 & 33.8 & 1.7 & 2.1 & G025.160+00.059 \\ 
 &  & &8.3 & 0.4 & 104.5 & 0.6 & 29.0 & 1.6 & 2.1 &  \\ 
25.308 & 0.494 & 2.3&5.2 & 0.3 & 40.4 & 0.8 & 24.3 & 1.8 & 2.1 & G025.305+00.532 \\ 
25.387 & 0.521 & 2.5&5.6 & 0.4 & 43.0 & 0.6 & 18.8 & 1.5 & 2.4 & G025.398+00.562 \\ 
25.492 & $-$0.193 & 2.2&16.1 & 0.4 & 107.3 & 0.4 & 24.8 & 1.2 & 2.1 & G025.460$-$00.210 \\ 
 &  & &10.4 & 0.3 & 62.6 & 0.5 & 33.2 & 1.4 & 2.1 &  \\ 
25.493 & $-$0.165 & 3.0&17.9 & 0.3 & 110.8 & 0.2 & 27.7 & 0.7 & 2.6 & G025.469$-$00.121 \\ 
 &  & &7.3 & 0.4 & 68.9 & 0.9 & 26.9 & 2.8 & 2.6 &  \\ 
25.495 & $-$0.001 & 2.7&12.1 & 0.4 & 56.4 & 0.7 & 27.4 & 1.5 & 1.9 & G025.477+00.040 \\ 
 &  & &5.8 & 0.4 & 100.5 & 1.7 & 41.4 & 5.1 & 1.9 &  \\ 
25.505 & $-$0.075 & 3.5&13.8 & 0.6 & 60.8 & 0.6 & 29.2 & 1.5 & 2.7 & G025.469$-$00.121 \\ 
 &  & &10.9 & 0.6 & 107.3 & 0.7 & 25.9 & 1.8 & 2.7 &  \\ 
25.659 & 0.065 & 3.2&14.7 & 0.7 & 120.0 & 0.3 & 11.2 & 0.6 & 2.8 & G025.700+00.030 \\ 
 &  & &8.5 & 0.4 & 89.5 & 0.8 & 32.9 & 2.3 & 2.8 &  \\ 
 &  & &7.1 & 0.5 & 48.3 & 0.6 & 15.3 & 1.4 & 2.8 &  \\ 
26.033 & $-$0.073 & 4.1&15.0 & 0.3 & 107.0 & 0.3 & 31.8 & 0.8 & 2.0 & G026.100$-$00.071 \\ 
 &  & &6.1 & 0.2 & 24.6 & 1.3 & 56.3 & 3.6 & 2.0 &  \\ 
26.235 & 0.241 & 2.8&13.6 & 0.5 & 108.0 & 0.7 & 35.5 & 1.9 & 3.0 & G026.261+00.280 \\ 
26.291 & 0.242 & 2.9&9.1 & 0.3 & 95.0 & 0.9 & 22.7 & 1.9 & 2.2 & G026.261+00.280 \\ 
 &  & &8.2 & 0.5 & 117.7 & 0.9 & 17.8 & 1.6 & 2.2 &  \\ 
26.557 & $-$0.036 & 2.7&8.1 & 0.3 & 103.1 & 0.9 & 45.3 & 2.1 & 2.3 & G026.599$-$00.019 \\ 
26.560 & $-$0.197 & 3.1&11.6 & 0.4 & 102.5 & 0.9 & 26.2 & 2.5 & 2.2 & G026.610$-$00.212 \\ 
 &  & &9.4 & 0.4 & 78.3 & 1.4 & 41.5 & 6.6 & 2.2 &  \\ 
26.678 & 0.003 & 2.3&17.1 & 0.6 & 69.2 & 0.3 & 15.2 & 0.7 & 2.5 & G026.644+00.020 \\ 
 &  & &13.7 & 0.4 & 101.3 & 0.6 & 34.5 & 1.5 & 2.5 &  \\ 
 &  & &7.0 & 0.5 & 26.6 & 0.6 & 16.5 & 1.4 & 2.5 &  \\ 
26.702 & 0.123 & 3.0&7.5 & 0.5 & 103.3 & 1.0 & 30.3 & 3.1 & 2.0 & G026.721+00.170 \\ 
 &  & &6.2 & 0.3 & 33.1 & 0.8 & 31.1 & 2.2 & 2.0 &  \\ 
 &  & &5.0 & 0.4 & 69.4 & 1.8 & 37.8 & 7.3 & 2.0 &  \\ 
27.217 & $-$0.064 & 1.9&11.7 & 0.3 & 91.6 & 0.4 & 32.1 & 1.1 & 2.4 & G027.189$-$00.079 \\ 
27.221 & $-$0.155 & 2.3&12.9 & 0.5 & 28.2 & 0.3 & 14.2 & 0.7 & 3.0 & G027.183$-$00.151 \\ 
 &  & &10.3 & 0.3 & 86.6 & 0.6 & 33.6 & 1.5 & 3.0 &  \\ 
27.333 & 0.131 & 2.7&14.0 & 0.4 & 96.7 & 0.4 & 26.3 & 1.2 & 2.3 & G027.334+00.176 \\ 
27.369 & 0.234 & 3.4&5.1 & 0.4 & 84.7 & 1.7 & 37.8 & 4.1 & 3.0 & G027.320+00.263 \\ 
27.371 & $-$0.125 & 2.3&17.6 & 0.4 & 94.1 & 0.9 & 28.2 & 1.7 & 2.1 & G027.366$-$00.164 \\ 
 &  & &7.4 & 0.4 & 73.0 & 0.6 & 21.5 & 1.6 & 2.1 &  \\ 
27.950 & $-$0.005 & 2.7&4.0 & 0.2 & 95.3 & 1.3 & 44.1 & 3.8 & 1.6 & G027.906$-$00.010 \\ 
 &  & &2.9 & 0.3 & 40.6 & 1.2 & 21.5 & 3.0 & 1.6 &  \\ 
27.962 & 0.198 & 1.7&10.0 & 0.3 & 92.6 & 0.4 & 26.8 & 1.1 & 2.3 & G027.934+00.206 \\ 
 &  & &6.4 & 0.5 & 38.2 & 0.4 & 11.2 & 1.0 & 2.3 &  \\ 
28.260 & $-$0.027 & 3.0&18.8 & 0.4 & 77.2 & 0.4 & 23.1 & 1.0 & 2.4 & G028.291+00.012 \\ 
 &  & &12.2 & 0.4 & 103.1 & 0.6 & 20.0 & 1.3 & 2.4 &  \\ 
 &  & &7.4 & 0.6 & 42.5 & 0.4 & 9.0 & 0.9 & 2.4 &  \\ 
28.282 & 0.059 & 2.9&14.5 & 0.3 & 93.4 & 0.4 & 41.3 & 1.2 & 2.3 & G028.291+00.012 \\ 
28.351 & 0.085 & 2.6&29.5 & 0.4 & 91.5 & 0.2 & 30.7 & 0.6 & 2.3 & G028.394+00.076 \\ 
28.382 & 0.035 & 2.6&24.9 & 0.3 & 89.0 & 0.2 & 28.1 & 0.4 & 2.0 & G028.394+00.076 \\ 
 &  & &6.2 & 0.4 & 45.7 & 0.5 & 16.6 & 1.2 & 2.0 &  \\ 
28.582 & 0.105 & 2.4&7.2 & 0.4 & 24.6 & 0.7 & 26.9 & 1.7 & 2.2 & G028.581+00.145 \\ 
 &  & &6.4 & 0.3 & 91.4 & 0.8 & 31.6 & 1.9 & 2.2 &  \\ 
28.644 & $-$0.045 & 4.5&13.9 & 0.5 & 97.4 & 0.4 & 22.6 & 1.1 & 2.5 & G028.609+00.021 \\ 
28.648 & 0.429 & 3.1&9.7 & 0.5 & 94.7 & 0.5 & 22.2 & 1.2 & 2.4 & G028.638+00.479 \\ 
28.737 & 0.068 & 2.7&8.8 & 0.5 & 98.7 & 0.6 & 20.2 & 1.4 & 2.5 & G028.696+00.048 \\ 
 &  & &7.3 & 0.4 & 46.4 & 0.9 & 30.4 & 2.7 & 2.5 &  \\ 
28.798 & 0.328 & 3.5&30.9 & 0.5 & 81.8 & 0.2 & 22.1 & 0.5 & 2.5 & G028.764+00.281 \\ 
 &  & &14.3 & 0.4 & 101.6 & 0.5 & 17.5 & 1.3 & 2.5 &  \\ 
29.180 & $-$0.139 & 2.8&4.4 & 0.3 & 58.6 & 1.7 & 41.7 & 6.4 & 2.0 & G029.134$-$00.145 \\ 
 &  & &3.7 & 0.5 & 97.9 & 1.3 & 19.3 & 3.1 & 2.0 &  \\ 
29.390 & 0.300 & 3.1&5.0 & 0.3 & 95.3 & 0.7 & 20.7 & 1.6 & 1.8 & G029.433+00.273 \\ 
 &  & &4.1 & 0.3 & 27.8 & 1.3 & 36.5 & 3.3 & 1.8 &  \\ 
29.401 & $-$0.376 & 4.7&4.1 & 0.3 & 96.0 & 1.4 & 42.5 & 3.8 & 2.4 & G029.344$-$00.430 \\ 
29.408 & 0.240 & 2.5&8.0 & 0.3 & 40.3 & 0.6 & 30.3 & 1.5 & 2.3 & G029.433+00.273 \\ 
 &  & &7.7 & 0.4 & 100.2 & 0.5 & 19.5 & 1.2 & 2.3 &  \\ 
29.774 & 0.175 & 2.6&5.4 & 0.5 & 87.9 & 1.4 & 29.8 & 3.7 & 2.6 & G029.770+00.219 \\ 
30.260 & 0.195 & 2.9&7.6 & 0.4 & 94.1 & 1.2 & 38.9 & 3.3 & 2.2 & G030.249+00.243 \\ 
30.293 & 0.030 & 2.8&46.4 & 0.8 & 104.3 & 0.2 & 22.4 & 0.4 & 2.6 & G030.252+00.051 \\ 
30.399 & 0.056 & 2.4&22.0 & 0.4 & 103.3 & 0.2 & 26.6 & 0.7 & 2.2 & G030.345+00.092, G030.374+00.026, G030.378+00.106 \\ 
30.402 & 0.137 & 2.3&25.7 & 0.5 & 109.9 & 0.2 & 18.3 & 0.5 & 2.1 & G030.378+00.106 \\ 
 &  & &9.7 & 0.5 & 67.4 & 0.7 & 21.1 & 2.1 & 2.1 &  \\ 
 &  & &8.8 & 0.6 & 88.5 & 0.6 & 12.6 & 1.6 & 2.1 &  \\ 
30.425 & $-$0.038 & 2.7&48.4 & 1.2 & 107.9 & 0.1 & 10.0 & 0.2 & 2.8 & G030.470$-$00.040 \\ 
 &  & &26.7 & 0.5 & 90.6 & 0.4 & 23.7 & 1.3 & 2.8 &  \\ 
30.442 & $-$0.165 & 2.0&22.3 & 0.8 & 81.3 & 0.3 & 17.1 & 0.9 & 2.3 & G030.463$-$00.191 \\ 
 &  & &16.2 & 0.3 & 107.6 & 0.6 & 44.1 & 3.3 & 2.3 &  \\ 
30.526 & $-$0.034 & 3.4&26.5 & 0.3 & 98.7 & 0.3 & 40.1 & 0.8 & 2.3 & G030.470$-$00.040 \\ 
 &  & &8.8 & 0.4 & 54.3 & 0.7 & 24.0 & 1.7 & 2.3 &  \\ 
30.629 & 0.015 & 2.1&19.9 & 0.4 & 102.3 & 0.3 & 33.0 & 0.9 & 2.2 & G030.597+00.030 \\ 
 &  & &7.3 & 0.4 & 40.9 & 0.7 & 24.3 & 1.7 & 2.2 &  \\ 
30.658 & 0.315 & 4.2&12.6 & 0.5 & 111.5 & 0.3 & 17.6 & 0.8 & 2.5 & G030.696+00.256 \\ 
 &  & &10.9 & 0.4 & 71.7 & 0.5 & 28.2 & 1.3 & 2.5 &  \\ 
30.720 & 0.225 & 2.7&25.8 & 0.4 & 98.5 & 0.2 & 28.6 & 0.5 & 1.9 & G030.760+00.206 \\ 
 &  & &7.7 & 0.5 & 38.5 & 0.5 & 17.8 & 1.2 & 1.9 &  \\ 
30.720 & $-$0.210 & 3.1&13.4 & 0.4 & 95.4 & 0.4 & 26.6 & 1.1 & 3.5 & G030.771$-$00.223 \\ 
30.734 & $-$0.605 & 2.6&10.2 & 0.3 & 94.5 & 0.5 & 39.6 & 1.4 & 1.9 & G030.698$-$00.628 \\ 
30.759 & 0.175 & 1.9&11.6 & 0.4 & 88.7 & 0.5 & 17.4 & 1.2 & 2.4 & G030.760+00.206 \\ 
 &  & &6.6 & 0.6 & 112.6 & 2.5 & 17.4 & 6.5 & 2.4 &  \\ 
 &  & &4.7 & 0.3 & 34.2 & 0.9 & 27.8 & 2.2 & 2.4 &  \\ 
30.777 & 0.247 & 2.7&22.2 & 0.4 & 35.6 & 0.2 & 20.2 & 0.5 & 2.3 & G030.760+00.206, G030.797+00.165, G030.838+00.114, G030.852+00.149 \\ 
 &  & &8.3 & 0.4 & 104.1 & 1.0 & 41.5 & 3.1 & 2.3 &  \\ 
30.823 & $-$0.282 & 4.7&22.7 & 0.5 & 92.4 & 0.3 & 25.3 & 0.7 & 3.4 & G030.771$-$00.223 \\ 
30.903 & 0.127 & 2.2&41.4 & 0.6 & 103.9 & 0.2 & 34.2 & 0.6 & 1.8 & G030.902+00.164 \\ 
 &  & &6.7 & 0.3 & 36.4 & 0.6 & 30.2 & 1.6 & 1.8 &  \\ 
30.915 & 0.054 & 2.2&53.1 & 0.6 & 99.2 & 0.2 & 30.9 & 0.4 & 2.6 & G030.883+00.071 \\ 
 &  & &16.2 & 0.8 & 38.1 & 0.5 & 19.5 & 1.1 & 2.6 &  \\ 
30.921 & 0.572 & 2.6&5.9 & 0.3 & 94.1 & 0.9 & 33.2 & 2.4 & 2.2 & G030.951+00.541, G030.956+00.599 \\ 
31.020 & 0.117 & 2.7&42.5 & 0.6 & 102.7 & 0.4 & 20.9 & 0.7 & 2.3 & G031.054+00.088 \\ 
 &  & &21.0 & 1.1 & 84.0 & 0.5 & 15.1 & 0.9 & 2.3 &  \\ 
 &  & &18.1 & 0.4 & 38.6 & 0.4 & 32.8 & 1.0 & 2.3 &  \\ 
31.078 & 0.175 & 3.4&25.5 & 0.4 & 108.2 & 0.2 & 19.5 & 0.5 & 2.1 & G031.104+00.125 \\ 
 &  & &22.6 & 0.4 & 84.0 & 0.2 & 18.1 & 0.6 & 2.1 &  \\ 
 &  & &11.2 & 0.3 & 43.2 & 0.4 & 24.2 & 0.9 & 2.1 &  \\ 
31.093 & 0.000 & 3.3&36.9 & 0.5 & 114.2 & 0.3 & 21.2 & 0.6 & 2.7 & G031.070+00.050 \\ 
 &  & &33.7 & 0.6 & 90.5 & 0.3 & 18.7 & 0.5 & 2.7 &  \\ 
 &  & &20.1 & 0.4 & 34.1 & 0.3 & 35.5 & 0.8 & 2.7 &  \\ 
31.109 & $-$0.151 & 2.9&64.1 & 0.5 & 43.7 & 0.1 & 18.1 & 0.2 & 3.3 & G031.157$-$00.148 \\ 
 &  & &7.0 & 0.6 & 20.1 & 0.8 & 15.1 & 2.2 & 3.3 &  \\ 
31.110 & 0.055 & 2.4&29.1 & 0.3 & 34.7 & 0.2 & 31.3 & 0.4 & 2.4 & G031.070+00.050 \\ 
 &  & &26.4 & 0.8 & 84.1 & 0.1 & 7.6 & 0.3 & 2.4 &  \\ 
 &  & &26.1 & 0.3 & 99.4 & 0.3 & 32.7 & 0.6 & 2.4 &  \\ 
31.128 & 0.084 & 2.3&26.4 & 0.4 & 104.3 & 0.4 & 25.5 & 0.9 & 2.9 & G031.104+00.125, G031.140+00.121, G031.159+00.048 \\ 
 &  & &19.9 & 0.9 & 83.9 & 0.3 & 13.1 & 0.6 & 2.9 &  \\ 
 &  & &12.6 & 0.4 & 40.9 & 0.6 & 21.3 & 1.4 & 2.9 &  \\ 
31.129 & 0.168 & 3.0&15.3 & 0.4 & 100.5 & 1.0 & 23.5 & 1.9 & 2.0 & G031.104+00.125 \\ 
 &  & &10.9 & 1.5 & 83.3 & 0.6 & 13.9 & 1.3 & 2.0 &  \\ 
 &  & &6.7 & 0.3 & 45.4 & 0.8 & 35.1 & 2.7 & 2.0 &  \\ 
31.167 & $-$0.177 & 1.9&10.9 & 0.4 & 44.2 & 0.3 & 18.7 & 0.8 & 2.2 & G031.157$-$00.148 \\ 
 &  & &9.7 & 0.3 & 94.2 & 0.5 & 29.6 & 1.2 & 2.2 &  \\ 
31.172 & 0.097 & 2.1&21.6 & 0.6 & 101.2 & 0.3 & 26.5 & 0.9 & 2.5 & G031.207+00.098 \\ 
 &  & &14.1 & 0.6 & 43.1 & 0.4 & 16.6 & 1.0 & 2.5 &  \\ 
31.198 & 0.133 & 2.2&16.7 & 0.4 & 100.8 & 0.4 & 30.9 & 1.2 & 2.0 & G031.207+00.098 \\ 
 &  & &8.0 & 0.4 & 43.0 & 0.9 & 35.1 & 2.4 & 2.0 &  \\ 
31.207 & 0.061 & 2.2&29.0 & 0.5 & 100.8 & 0.2 & 22.4 & 0.5 & 2.8 & G031.207+00.098 \\ 
 &  & &8.7 & 0.4 & 43.0 & 0.4 & 18.8 & 1.0 & 2.8 &  \\ 
31.232 & $-$0.065 & 2.8&16.6 & 0.4 & 99.3 & 0.4 & 21.9 & 1.0 & 2.2 & G031.240$-$00.110 \\ 
 &  & &16.2 & 0.3 & 34.4 & 0.4 & 42.5 & 1.2 & 2.2 &  \\ 
 &  & &7.5 & 0.6 & 77.2 & 0.6 & 12.8 & 1.5 & 2.2 &  \\ 
31.233 & $-$0.070 & 2.4&17.0 & 0.3 & 33.7 & 0.4 & 45.8 & 1.3 & 3.0 & G031.240$-$00.110 \\ 
 &  & &15.2 & 0.5 & 98.7 & 0.4 & 21.4 & 1.1 & 3.0 &  \\ 
 &  & &9.2 & 0.8 & 79.8 & 0.4 & 9.9 & 1.1 & 3.0 &  \\ 
31.416 & $-$0.106 & 2.2&18.2 & 0.6 & 97.1 & 0.3 & 18.8 & 0.8 & 2.3 & G031.425$-$00.141 \\ 
 &  & &6.2 & 0.4 & 80.0 & 3.9 & 43.7 & 8.4 & 2.3 &  \\ 
31.517 & $-$0.068 & 2.9&7.3 & 0.6 & 98.3 & 0.7 & 15.7 & 1.6 & 2.5 & G031.535$-$00.113 \\ 
31.566 & 0.270 & 4.5&5.7 & 0.4 & 116.1 & 0.5 & 12.8 & 1.3 & 2.1 & G031.607+00.334 \\ 
 &  & &5.1 & 0.3 & 87.4 & 1.0 & 27.0 & 3.4 & 2.1 &  \\ 
31.631 & $-$0.291 & 4.0&9.8 & 0.3 & 96.6 & 0.4 & 23.2 & 1.0 & 2.3 & G031.600$-$00.232 \\ 
31.760 & $-$0.112 & 3.4&35.1 & 0.6 & 94.8 & 0.1 & 12.7 & 0.3 & 2.1 & G031.816$-$00.121 \\ 
 &  & &10.3 & 0.6 & 42.2 & 0.4 & 16.1 & 1.0 & 2.1 &  \\ 
 &  & &9.5 & 0.8 & 110.1 & 0.4 & 8.5 & 1.0 & 2.1 &  \\ 
32.057 & 0.007 & 4.0&6.4 & 0.5 & 91.4 & 0.8 & 20.7 & 2.0 & 2.9 & G032.108+00.050 \\ 
32.155 & 0.011 & 3.7&23.7 & 0.6 & 41.8 & 0.2 & 13.6 & 0.4 & 1.9 & G032.108+00.050 \\ 
 &  & &10.5 & 0.6 & 95.1 & 0.3 & 11.6 & 0.8 & 1.9 &  \\ 
32.552 & $-$0.048 & 2.4&7.2 & 0.6 & 27.8 & 0.6 & 13.6 & 1.4 & 2.4 & G032.581$-$00.077 \\ 
 &  & &5.0 & 0.5 & 101.4 & 1.0 & 17.8 & 2.3 & 2.4 &  \\ 
33.790 & $-$0.088 & 2.7&6.0 & 0.4 & 48.4 & 1.4 & 47.8 & 4.1 & 2.2 & G033.753$-$00.063 \\ 
 &  & &5.6 & 0.5 & 102.0 & 1.1 & 22.5 & 3.0 & 2.2 &  \\ 
33.965 & 0.025 & 2.5&5.6 & 0.3 & 43.6 & 1.0 & 34.4 & 2.1 & 2.0 & G033.994$-$00.006 \\ 
34.065 & $-$0.092 & 2.9&10.6 & 0.3 & 40.0 & 0.3 & 21.2 & 0.7 & 2.2 & G034.031$-$00.059 \\ 
34.587 & 0.203 & 2.5&10.0 & 0.5 & 55.6 & 0.4 & 22.8 & 1.0 & 2.2 & G034.591+00.244 \\ 
34.750 & $-$0.041 & 3.7&7.9 & 0.3 & 52.8 & 0.5 & 28.6 & 1.3 & 2.1 & G034.750+00.020 \\ 
34.765 & 0.074 & 3.4&17.9 & 0.5 & 83.1 & 0.2 & 13.8 & 0.5 & 2.2 & G034.750+00.020 \\ 
 &  & &14.6 & 0.4 & 53.2 & 0.3 & 19.5 & 0.7 & 2.2 &  \\ 
34.822 & 0.024 & 4.3&9.5 & 0.3 & 50.6 & 0.5 & 27.1 & 1.7 & 2.2 & G034.750+00.020 \\ 
 &  & &8.2 & 0.4 & 83.5 & 0.5 & 17.5 & 1.2 & 2.2 &  \\ 
35.220 & 0.093 & 3.0&3.9 & 0.3 & 56.2 & 2.2 & 38.7 & 6.5 & 1.8 & G035.260+00.122 \\ 
35.305 & 0.104 & 2.9&6.2 & 0.5 & 46.3 & 0.5 & 14.3 & 1.3 & 1.8 & G035.260+00.122 \\ 
36.953 & 0.102 & 4.6&4.1 & 0.2 & 39.2 & 0.8 & 35.4 & 1.9 & 1.5 & G037.027+00.123 \\ 
37.459 & $-$0.409 & 4.6&6.0 & 0.4 & 47.7 & 0.3 & 9.5 & 0.8 & 2.1 & G037.432$-$00.337 \\ 
37.546 & 0.543 & 3.0&5.9 & 0.2 & 45.8 & 0.5 & 24.7 & 1.5 & 1.9 & G037.485+00.513, G037.498+00.530 \\ 
38.110 & $-$0.200 & 1.7&18.4 & 0.6 & 53.9 & 0.2 & 12.0 & 0.5 & 2.5 & G038.120$-$00.227 \\ 
 &  & &8.1 & 0.6 & 71.9 & 0.5 & 11.7 & 1.2 & 2.5 &  \\ 
50.057 & $-$0.143 & 3.7&6.4 & 0.3 & 64.7 & 0.3 & 11.9 & 0.7 & 1.9 & G049.998$-$00.125 \\ 

%% file: reasons.tab
1		&	\phn\phn \input reason1	&	\phn 4	& Single velocity on-target \\
2		&	\phn\phn\input reason4	&	\phn 4	& One component negative velocity\\
3a		&	\phn \input reason2	&	26	& 50mK decrease\\
3b		&	\phn\input reason3	&	28	& 20mK decrease\\
4a		&	\phn\input reason5	&	32	& High-quality \te\\
4b		&	\phn\input reason6	&	14	& Low-quality \te\\
5a		&	\phn\phn\input reason7	&	\phn 8	& CO morphology\\
5b		&	\phn\input reason8	&	28	& Other molecular lines\\
6		&	\phn\input reason9	&	\phn 5	& Carbon RRL\\
\hline
1 Criterion	&	\phn 47     		&	40	& \\
2 Criteria	&	\phn 39			&	33	& \\
$>2$ Criteria	&	\phn 17			&	15	& \\
Total Resolved	&	\input resolved 	&	88	& \\

%% file: reason4
5

%% file: reason2
31

%% file: reason3
33

%% file: reason5
37

%% file: reason6
17

%% file: reason7
9

%% file: reason8
33

%% file: reason9
6

%% file: distances_lda.tab
G018.677$-$00.236 & \phn \phs 42.6 & 3.5 & 12.6 & 8.1 & F & A & \phn 5.3 & 12.6 & $ -51$ & 3b,4a,5a\\
G019.594+00.024 & \phn \phs 35.3 & 3.0 & 13.0 & 8.0 & \nodata & C & \phn 5.7 & \nodata & \nodata & 4b\\
G019.728$-$00.113 & \phn \phs 57.3 & 4.2 & 11.8 & 8.0 & \nodata & C & \phn 4.8 & \nodata & \nodata & 5a\\
G019.813+00.010 & \phn \phs 60.4 & 4.3 & 11.7 & 8.0 & \nodata & C & \phn 4.7 & \nodata & \nodata & 4a\\
G021.558$-$00.112 & \phs 115.1 & 6.5 & \phn 9.3 & 7.9 & \nodata & C & \phn 3.4 & \nodata & \nodata & 1\\
G021.744$-$00.032 & \phn \phs 20.0 & 1.8 & 14.0 & 7.9 & \nodata & C & \phn 6.9 & \nodata & \nodata & 3b, 5a\\
G021.933+00.103 & \phn \phs 82.2 & 5.1 & 10.6 & 7.9 & \nodata & C & \phn 4.2 & \nodata & \nodata & 4b,5a\\
G022.730$-$00.239 & \phn \phs 71.1 & 4.6 & 11.0 & 7.8 & \nodata & C & \phn 4.6 & \nodata & \nodata & 3b,4a\\
G022.755$-$00.246 & \phs 106.7 & 6.1 & \phn 9.6 & 7.8 & N & B & \phn 3.7 & 6.1 & $ -26$ & 5a\\
G023.200+00.000 & \phn \phs 89.5 & 5.4 & 10.2 & 7.8 & \nodata & C & \phn 4.1 & \nodata & \nodata & 1\\
G023.265$-$00.301 & \phn \phs 73.1 & 4.7 & 10.9 & 7.8 & N & A & \phn 4.6 & 4.7 & $ -24$ & 3a\\
G023.389$-$00.148 & \phn \phs 98.5 & 5.8 & \phn 9.8 & 7.8 & N & A & \phn 3.9 & 5.8 & $ -14$ & 3a,4a\\
G023.458$-$00.179 & \phn \phs 97.4 & 5.7 & \phn 9.9 & 7.8 & N & A & \phn 4.0 & 5.7 & $ -17$ & 3a,4b,5a\\
G023.513$-$00.244 & \phn \phs 81.4 & 5.1 & 10.5 & 7.8 & \nodata & C & \phn 4.4 & \nodata & \nodata & 3b\\
G023.604$-$00.200 & \phn \phs 82.7 & 5.1 & 10.5 & 7.8 & \nodata & C & \phn 4.3 & \nodata & \nodata & 3b,4b\\
G023.736$-$00.022 & \phn \phn \phs 3.0 & 0.2 & 15.4 & 7.8 & F & B & \phn 8.3 & 15.4 & $  -5$ & 4a,5a\\
G023.836+00.104 & \phn \phs 41.9 & 3.1 & 12.4 & 7.8 & F & A & \phn 5.8 & 12.4 & $  22$ & 3b\\
G023.909+00.066 & \phn \phs 38.4 & 2.9 & 12.6 & 7.8 & F & A & \phn 5.9 & 12.6 & $  14$ & 3b,5b\\
G024.300$-$00.149 & \phn \phs 56.8 & 3.9 & 11.6 & 7.7 & \nodata & C & \phn 5.2 & \nodata & \nodata & 3a\\
G024.400$-$00.190 & \phn \phs 54.7 & 3.8 & 11.7 & 7.7 & \nodata & C & \phn 5.3 & \nodata & \nodata & 5b,5a\\
G024.500+00.487 & \phn \phs 98.9 & 5.8 & \phn 9.7 & 7.7 & N & B & \phn 4.0 & 5.8 & $  49$ & 3a,4b\\
G024.507$-$00.222 & \phn \phs 96.3 & 5.7 & \phn 9.8 & 7.7 & N & A & \phn 4.1 & 5.7 & $ -22$ & 3a,5b\\
G024.557$-$00.132 & \phn \phs 91.1 & 5.4 & 10.0 & 7.7 & \nodata & C & \phn 4.2 & \nodata & \nodata & 3b,4a,5a\\
G024.680$-$00.160 & \phs 110.3 & 6.4 & \phn 9.1 & 7.7 & N & B & \phn 3.8 & 6.4 & $ -17$ & 3a,5b,5a\\
G024.713$-$00.125 & \phs 112.3 & 6.5 & \phn 9.0 & 7.7 & N & B & \phn 3.8 & 6.5 & $ -14$ & 3a,4b,5a\\
G024.735+00.159 & \phs 109.3 & 6.3 & \phn 9.1 & 7.7 & \nodata & C & \phn 3.8 & \nodata & \nodata & 3a,5a\\
G024.744$-$00.206 & \phn \phs 82.6 & 5.1 & 10.4 & 7.7 & F & A & \phn 4.4 & 10.4 & $ -37$ & 3a\\
G024.923+00.079 & \phn \phs 42.4 & 3.1 & 12.3 & 7.7 & F & A & \phn 5.8 & 12.3 & $  16$ & 3b,4a,5a\\
G024.955$-$00.048 & \phn \phs 49.8 & 3.5 & 11.9 & 7.7 & F & B & \phn 5.5 & 11.9 & $  -9$ & 4a\\
G025.150+00.092 & \phn \phs 46.5 & 3.3 & 12.1 & 7.7 & F & B & \phn 5.7 & 12.1 & $  19$ & 3a,4a,5a\\
G025.160+00.059 & \phn \phs 45.5 & 3.3 & 12.1 & 7.7 & F & B & \phn 5.7 & 12.1 & $  12$ & 3a,5b\\
G025.305+00.532 & \phn \phn \phs 7.0 & 0.6 & 14.8 & 7.7 & \nodata & C & \phn 8.0 & \nodata & \nodata & 3b,4a\\
G025.398+00.562 & \phn \phs 12.9 & 1.1 & 14.3 & 7.7 & F & B & \phn 7.6 & 14.3 & $ 140$ & 3b,5a\\
G025.460$-$00.210 & \phs 117.3 & 6.9 & \phn 8.4 & 7.7 & T & \nodata & \phn 3.7 & 7.7 & $ -28$ & 3b,5b\\
G025.469$-$00.121 & \phn \phs 96.9 & 5.7 & \phn 9.6 & 7.7 & \nodata & C & \phn 4.1 & \nodata & \nodata & 3a,4a\\
G025.477+00.040 & \phn $-14.6$ & \nodata & 17.3 & 7.7 & F & A & 10.3 & 17.3 & $  12$ & 2\\
G025.700+00.030 & \phn \phs 52.7 & 3.6 & 11.7 & 7.7 & \nodata & C & \phn 5.5 & \nodata & \nodata & 3a\\
G026.100$-$00.071 & \phn \phs 30.7 & 2.3 & 12.9 & 7.6 & F & A & \phn 6.5 & 12.9 & $ -15$ & 3a\\
G026.261+00.280 & \phn \phs 88.6 & 5.3 & \phn 9.9 & 7.6 & \nodata & C & \phn 4.4 & \nodata & \nodata & 4a\\
G026.599$-$00.019 & \phn \phs 19.4 & 1.5 & 13.7 & 7.6 & \nodata & C & \phn 7.2 & \nodata & \nodata & 3b\\
G026.610$-$00.212 & \phn $-35.7$ & \nodata & 20.8 & 7.6 & F & A & 13.8 & 20.8 & $ -77$ & 3b,2,4a\\
G026.644+00.020 & \phs 108.7 & 6.4 & \phn 8.8 & 7.6 & T & \nodata & \phn 4.0 & 7.6 & $   2$ & 4a,5a\\
G026.721+00.170 & \phn \phs 31.0 & 2.3 & 12.9 & 7.6 & F & A & \phn 6.5 & 12.9 & $  38$ & 3b,4a,5a\\
G027.183$-$00.151 & \phn \phs 27.2 & 2.1 & 13.1 & 7.6 & F & B & \phn 6.7 & 13.1 & $ -34$ & 4a\\
G027.189$-$00.079 & \phn \phs 27.8 & 2.1 & 13.0 & 7.6 & F & B & \phn 6.7 & 13.0 & $ -17$ & 5a\\
G027.320+00.263 & \phn \phs 28.5 & 2.2 & 13.0 & 7.6 & \nodata & C & \phn 6.7 & \nodata & \nodata & 3b\\
G027.334+00.176 & \phn \phs 80.4 & 4.9 & 10.2 & 7.6 & F & B & \phn 4.7 & 10.2 & $  31$ & 3b,4a\\
G027.906$-$00.010 & \phn \phs 96.4 & 5.8 & \phn 9.2 & 7.5 & \nodata & C & \phn 4.3 & \nodata & \nodata & 4a\\
G027.934+00.206 & \phn \phs 50.6 & 3.4 & 11.6 & 7.5 & \nodata & C & \phn 5.7 & \nodata & \nodata & 4a,5a\\
G028.291+00.012 & \phn \phs 44.4 & 3.1 & 11.9 & 7.5 & \nodata & C & \phn 6.0 & \nodata & \nodata & 5a\\
G028.303$-$00.388 & \phn \phs 76.9 & 4.8 & 10.2 & 7.5 & \nodata & C & \phn 4.9 & \nodata & \nodata & 1\\
G028.394+00.076 & \phn \phs 86.2 & 5.2 & \phn 9.7 & 7.5 & \nodata & C & \phn 4.6 & \nodata & \nodata & 4a\\
G028.581+00.145 & \phn $-13.0$ & \nodata & 16.5 & 7.5 & F & A & \phn 9.9 & 16.5 & $  41$ & 2,4a\\
G028.609+00.021 & \phn \phs 96.6 & 5.8 & \phn 9.1 & 7.5 & F & A & \phn 4.4 & 9.1 & $   3$ & 3a\\
G028.638+00.479 & \phn \phs 28.4 & 2.1 & 12.8 & 7.5 & F & A & \phn 6.7 & 12.8 & $ 107$ & 3b,4a\\
G028.696+00.048 & \phs 100.7 & 6.1 & \phn 8.8 & 7.5 & F & B & \phn 4.3 & 8.8 & $   7$ & 3a,4a\\
G028.764+00.281 & \phs 104.5 & 6.4 & \phn 8.5 & 7.4 & T & \nodata & \phn 4.2 & 7.4 & $  36$ & 3b\\
G029.134$-$00.145 & \phn \phs 48.6 & 3.3 & 11.6 & 7.4 & \nodata & C & \phn 5.8 & \nodata & \nodata & 4b\\
G029.344$-$00.430 & \phn \phs 91.6 & 5.6 & \phn 9.2 & 7.4 & \nodata & C & \phn 4.6 & \nodata & \nodata & 4a\\
G029.770+00.219 & \phn $-31.3$ & \nodata & 18.9 & 7.4 & F & A & 12.3 & 18.9 & $  72$ & 2,4a\\
G029.981$-$00.607 & \phn \phs 90.9 & 5.6 & \phn 9.2 & 7.4 & F & B & \phn 4.6 & 9.2 & $ -97$ & 1\\
G030.249+00.243 & \phn \phn \phs 8.9 & 0.6 & 14.1 & 7.3 & F & A & \phn 8.0 & 14.1 & $  59$ & 3b,4a,6\\
G030.252+00.051 & \phn \phs 65.2 & 4.2 & 10.5 & 7.3 & F & B & \phn 5.3 & 10.5 & $   9$ & 3b,4a,5a\\
G030.345+00.092 & \phs 106.7 & 7.3 & \phn 7.3 & 7.3 & T & \nodata & \phn 4.3 & 7.3 & $  11$ & 4b\\
G030.374+00.026 & \phn \phs 44.4 & 3.0 & 11.6 & 7.3 & F & A & \phn 6.1 & 11.6 & $   5$ & 3a,4a,5a\\
G030.378+00.106 & \phs 102.0 & 6.5 & \phn 8.2 & 7.3 & T & \nodata & \phn 4.4 & 7.3 & $  13$ & 5a\\
G030.470$-$00.040 & \phn \phs 42.7 & 2.9 & 11.7 & 7.3 & F & A & \phn 6.2 & 11.7 & $  -8$ & 3a\\
G030.597+00.030 & \phn \phs 44.9 & 3.0 & 11.6 & 7.3 & \nodata & C & \phn 6.1 & \nodata & \nodata & 3b,4b\\
G030.644+00.053 & \phn \phs 99.5 & 6.3 & \phn 8.4 & 7.3 & T & \nodata & \phn 4.5 & 7.3 & $   6$ & 3b,4a\\
G030.696+00.256 & \phs 102.1 & 6.6 & \phn 8.1 & 7.3 & T & \nodata & \phn 4.4 & 7.3 & $  32$ & 3a,4b\\
G030.760+00.206 & \phn \phs 98.5 & 6.2 & \phn 8.4 & 7.3 & T & \nodata & \phn 4.5 & 7.3 & $  26$ & 3b,5a\\
G030.771$-$00.223 & \phs 103.6 & 6.8 & \phn 7.8 & 7.3 & T & \nodata & \phn 4.4 & 7.3 & $ -28$ & 3a,4a\\
G030.797+00.165 & \phn \phs 40.4 & 2.8 & 11.8 & 7.3 & F & A & \phn 6.3 & 11.8 & $  34$ & 3a,4a\\
G030.838+00.114 & \phn \phs 35.8 & 2.5 & 12.1 & 7.3 & F & A & \phn 6.5 & 12.1 & $  24$ & 3a,4b,6\\
G030.852+00.149 & \phn \phs 39.6 & 2.7 & 11.9 & 7.3 & F & A & \phn 6.3 & 11.9 & $  30$ & 3a,4b\\
G030.883+00.071 & \phn \phs 96.5 & 6.0 & \phn 8.6 & 7.3 & T & \nodata & \phn 4.5 & 7.3 & $   9$ & 3b,6\\
G030.902+00.164 & \phn \phs 99.3 & 6.3 & \phn 8.3 & 7.3 & T & \nodata & \phn 4.5 & 7.3 & $  20$ & 5a\\
G030.951+00.541 & \phn \phs 24.5 & 1.8 & 12.8 & 7.3 & F & A & \phn 7.0 & 12.8 & $ 120$ & 3a,4a\\
G030.956+00.599 & \phn \phs 23.4 & 1.7 & 12.9 & 7.3 & F & A & \phn 7.1 & 12.9 & $ 134$ & 3a,6\\
G031.054+00.088 & \phn \phs 20.1 & 1.5 & 13.1 & 7.3 & F & A & \phn 7.3 & 13.1 & $  20$ & 3a,5b\\
G031.070+00.050 & \phn \phs 35.4 & 2.5 & 12.1 & 7.3 & F & A & \phn 6.5 & 12.1 & $  10$ & 3a\\
G031.140+00.121 & \phn \phs 45.5 & 3.1 & 11.5 & 7.3 & \nodata & C & \phn 6.1 & \nodata & \nodata & 3b\\
G031.157$-$00.148 & \phn \phs 43.6 & 3.0 & 11.6 & 7.3 & F & A & \phn 6.2 & 11.6 & $ -29$ & 3a,4a,5a,6\\
G031.159+00.048 & \phn \phs 35.6 & 2.5 & 12.1 & 7.3 & F & B & \phn 6.5 & 12.1 & $  10$ & 4a,5a\\
G031.240$-$00.110 & \phn \phs 26.5 & 1.9 & 12.6 & 7.3 & F & A & \phn 6.9 & 12.6 & $ -24$ & 3a,5b\\
G031.425$-$00.141 & \phn \phs 28.8 & 2.0 & 12.5 & 7.3 & \nodata & C & \phn 6.8 & \nodata & \nodata & 3b,5a\\
G031.600$-$00.232 & \phn \phs 40.8 & 2.8 & 11.7 & 7.2 & \nodata & C & \phn 6.3 & \nodata & \nodata & 3b,5b\\
G031.607+00.334 & \phn \phs 24.4 & 1.8 & 12.7 & 7.2 & F & A & \phn 7.1 & 12.7 & $  74$ & 3b\\
G031.816$-$00.121 & \phn \phs 36.3 & 2.5 & 11.9 & 7.2 & F & A & \phn 6.5 & 11.9 & $ -25$ & 3b\\
G032.108+00.050 & \phn \phs 94.4 & 6.0 & \phn 8.4 & 7.2 & T & \nodata & \phn 4.7 & 7.2 & $   6$ & 3b,4a\\
G032.379+00.102 & \phn \phs 47.3 & 3.2 & 11.2 & 7.2 & \nodata & C & \phn 6.1 & \nodata & \nodata & 1\\
G032.581$-$00.077 & \phn \phs 30.0 & 2.1 & 12.2 & 7.2 & \nodata & C & \phn 6.8 & \nodata & \nodata & 4b\\
G033.753$-$00.063 & \phs 101.7 & 7.1 & \phn 7.1 & 7.1 & T & \nodata & \phn 4.7 & 7.1 & $  -7$ & 4a\\
G034.031$-$00.059 & \phn \phs 56.3 & 3.7 & 10.4 & 7.0 & \nodata & C & \phn 5.8 & \nodata & \nodata & 3a,4b,6\\
G034.197$-$00.592 & \phn \phs 52.4 & 3.4 & 10.6 & 7.0 & \nodata & C & \phn 6.0 & \nodata & \nodata & 5a\\
G034.591+00.244 & \phn $-19.4$ & \nodata & 16.1 & 7.0 & F & A & 10.3 & 16.1 & $  68$ & 2\\
G034.750+00.020 & \phn \phs 86.1 & 5.7 & \phn 8.3 & 7.0 & T & \nodata & \phn 5.0 & 7.0 & $   2$ & 3b,5a\\
G035.260+00.122 & \phn \phs 38.1 & 2.6 & 11.3 & 6.9 & \nodata & C & \phn 6.6 & \nodata & \nodata & 4b\\
G037.432$-$00.337 & \phn \phs 44.1 & 2.9 & 10.6 & 6.8 & \nodata & C & \phn 6.4 & \nodata & \nodata & 4b\\
G037.485+00.513 & \phn \phn \phs 4.1 & 0.2 & 13.3 & 6.7 & F & B & \phn 8.4 & 13.3 & $ 119$ & 4b\\
G037.498+00.530 & \phn \phs 11.6 & 0.8 & 12.7 & 6.7 & \nodata & C & \phn 7.9 & \nodata & \nodata & 5a\\
G038.120$-$00.227 & \phn \phs 83.8 & 6.7 & \phn 6.7 & 6.7 & T & \nodata & \phn 5.2 & 6.7 & $ -26$ & 5a\\
G049.998$-$00.125 & \phn \phs 38.5 & 3.0 & \phn 7.9 & 5.5 & F & B & \phn 7.0 & 7.9 & $ -17$ & 3b,4a,5a\\